\documentclass[twocolumn,pre,showpacs,eqsecnum]{revtex4}

\usepackage{dcolumn}
\usepackage{amsfonts}
\usepackage{amsmath}
\usepackage{amssymb}
\usepackage{bm}
\usepackage{epsfig}

\begin{document}

\newcommand{\brm}[1]{\bm{{\rm #1}}}
\newcommand{\tens}[1]{\underline{\underline{#1}}}
\newcommand{\mm}{\overset{\leftrightarrow}{m}}
\newcommand{\xv}{\bm{{\rm x}}}
\newcommand{\Rv}{\bm{{\rm R}}}
\newcommand{\uv}{\bm{{\rm u}}}
\newcommand{\nv}{\bm{{\rm n}}}
\newcommand{\Nv}{\bm{{\rm N}}}
\newcommand{\ev}{\bm{{\rm e}}}

\title{Dynamics of smectic  elastomers}

\author{Olaf Stenull}
\affiliation{Fachbereich Physik, Universit\"{a}t Duisburg-Essen, Campus Duisburg, 47048 Duisburg,
Germany }

\author{T. C. Lubensky}
\affiliation{Department of Physics and Astronomy, University of
Pennsylvania, Philadelphia, PA 19104, USA }

\vspace{10mm}
\date{\today}

\begin{abstract}
We study the low-frequency, long-wavelength dynamics of liquid crystal elastomers, crosslinked in the smectic-$A$ phase, in their smectic-$A$, biaxial smectic and smectic-$C$ phases. Two different yet related formulations are employed. One formulation describes the pure hydrodynamics and does not explicitly involve the Frank director, which relaxes to its local equilibrium value in a non-hydrodynamic time. The other formulation explicitly treats the director and applies beyond the hydrodynamic limit. We compare the low-frequency, long-wavelength dynamics of smectic-$A$ elastomers to that of nematics and show that the two are closely related. For the biaxial smectic and the smectic-$C$ phases, we calculate sound velocities and the mode structure in certain symmetry directions. For the smectic-$C$ elastomers, in addition, we discuss in some detail their possible behavior in rheology experiments.
\end{abstract}


\maketitle

\section{introduction}
Liquid crystal elastomers~\cite{WarnerTer2003} are soft amorphous solids that have the macroscopic symmetry properties of liquid crystals~\cite{deGennesProst93_Chandrasekhar92}. Usually, they consist of crosslinked side-chain or main-chain liquid crystal polymers. In their smectic phases, these elastomers
possess a plane-like, lamellar modulation of density in one direction. In the smectic-$A$ (Sm$A$) phase, the director describing the average orientation of constituent mesogens is parallel to the normal to the smectic layers whereas in the smectic-$C$ (Sm$C$) phase, it has a component in the plane of the
layers. 

Various unusual properties of smectic elastomers have been discovered to date experimentally and/or theoretically, some of which are truly remarkable. For example, in smectic elastomers the crosslinking can suppress the Landau-Peierls-Instability~\cite{landau_65,peierls_34,terentjev_warner_lubensky_95} that leads to a breakdown of one-dimensional long-range order in other layered systems such as stacked surfactant membranes~\cite{safinia&Co_86,sirota&Co_88} or conventional smectic liquid crystals~\cite{als-nielsen&Co_80}. Depending on the chemical composition of the material~\cite{nishikawa_finkelmann_98}, the stabilizing effect of the elastic network can be strong enough to allow for the persistence of one-dimensional long-range order over several microns~\cite{wong&Co_97}.  Sm$A$ elastomers exhibit a non-Hookean elasticity when stretched along the normal of the smectic layers with a high Young's modulus $Y_\parallel$ for strains below a threshold of about $3\%$ and a considerably smaller $Y_\parallel$ for strains above that threshold~\cite{nishikawa_finkelmann_99,adams_warner_SmA_2005,stenull_lubensky_SmA_2006}. Sm$C$ elastomers prepared by crosslinking in the Sm$A$ phase followed by a cooling into the Sm$C$ phase, like nematic elastomers~\cite{golubovic_lubensky_89,FinKun97,VerWar96,War99,LubenskyXin2002}, are predicted to exhibit the phenomenon of {\em soft elasticity}~\cite{stenull_lubensky_SmCsoftness_2005, stenull_lubensky_SmCsoftness_2006,adams_warner_SmCsoftness_2005}. The phase transition from the Sm$A$ to the Sm$C$ phase is associated with a spontaneous breaking of the rotational invariance in the smectic plane. The Goldstone mode of this symmetry breaking is manifests  itself in the vanishing of certain shear moduli as well as the vanishing of the energy cost of certain nonlinear extensional strains.

It is the interplay between elastic and liquid crystalline degrees of freedom that leads to these and other phenomena that are neither present in conventional rubber nor in conventional liquid crystals. The effects of this interplay are strongest in ideal, monodomain samples of liquid crystal elastomers, with a homogeneous orientation of constituent mesogens. Generically, liquid crystal elastomers tend to segregate into many domains, each having its own local Frank director $\brm{n}$ specifying the direction of mesogenic order. In order to avoid such polydomain samples, elaborate synthetization methods have been developed and applied to smectic elastomers for more than ten years now~\cite{bremer&Co_93,bremer&Co_94,benne&Co_94}. Most fruitful, perhaps, have been multistage crosslinking techniques that allow for monodomain samples that neither need to be confined between director-aligning plates nor need to be kept in director-aligning external fields. These techniques have been used, for example, to synthesize elastomers with a permanent, macroscopically ordered Sm$A$ monodomain structure~\cite{nishikawa_finkelmann_97} as well as monodomain Sm$C$ elastomers that exhibit reversible phase transitions to the Sm$A$ and isotropic phases in response to temperature change~\cite{hiraoka_finkelmann_2001,hiraoka&Co_2005}.

Despite the successes in understanding the static properties of smectic elastomers and the advances in their synthesis, dynamical experiments on these materials, such as rheology experiments of storage and loss moduli or Brillouin scattering measurements of sound velocities, have not been reported to date. To interpret these kinds of experiments once they are done, it is desirable to have theories for the low-frequency, long-wavelength dynamics of smectic elastomers. Recently, we introduced such a theory for Sm$C$ elastomers~\cite{stenull_lubensky_SmCdynamics_2006}. Here, we present this work in more detail and we extend it to the Sm$A$ and biaxial smectic phases. To keep our discussion as simple as possible, we will exclusively consider elastomers, that have been crosslinked in the Sm$A$ phase, so that the smectic layers are locked to the crosslinked network~\cite{lubensky&Co_94,massDensWaves}. To be specific, we will consider Sm$A$ elastomers crosslinked in the Sm$A$ phase and soft Sm$C$ elastomers that form spontaneously from these Sm$A$ elastomers when temperature falls below the Sm$A$-to-Sm$C$ transition temperature. In addition to these materials, we study soft biaxial smectic elastomers that also could form spontaneously from a Sm$A$ elastomer, at least in principle, but that may be, as biaxial phases in general, hard to find in nature. We include these elastomers in our study mainly because their dynamics is similar to but simpler than that of soft Sm$C$ elastomers and, therefore, has a potential pedagogical value. 

Our theories fall into two categories. The first category is pure hydrodynamics which describes the dynamics of those degrees of freedom whose characteristic frequencies $\omega$ vanish as wavenumber $\brm{q}$ goes to zero. These theories focus on the very leading low-frequency, long-wavelength behavior and apply for frequencies $\omega$ such that $\omega \, \tau \ll 1$, where $\tau$ is the longest non-hydrodynamic decay time in the system. As in conventional elastic media, our hydrodynamic theories involve only the elastic displacement field $\brm{u}$ and not the director $\brm{n}$, which relaxes to the local strain in a non-hydrodynamic time $\tau_n$. The second category of theories explicitly includes $\brm{n}$ and applies for a larger range of frequencies than pure hydrodynamics. One motivation for setting up these theories is to better understand the role of $\brm{n}$ in the dynamics of smectic elastomers. The other motivation is that dynamical experiments, like rheology measurements, typically probe a wide range of frequencies that extends from the hydrodynamic regime to frequencies well above it, and that, therefore, theories going beyond hydrodynamics could be valuable for interpreting these experiments. Our theories with displacement and director assume that $\tau_n$ is the longest non-hydrodynamic relaxation time, $\tau = \tau_n$, and that $\tau_n$ is well separated from the next longest non-hydrodynamic time $\tau_E$, which we refer to as elastomer time. The specific origin of $\tau_E$ is not essential to our work. For example, $\tau_E$ might stem from Rouse-like dynamics of the polymers constituting the elastomeric matrix. Provided the assumption $\tau_n \gg \tau_E$ holds, our dynamics with $\brm{u}$ and $\brm{n}$ applies for $\omega \, \tau_n \ll 1$ and  $\omega \, \tau_n > 1$.

The outline of the remainder of this paper is as follows. Section~\ref{dynamicalEquationsReview} briefly reviews the well established Poisson bracket formalism for coarse grained variables as well as the resulting equations of motion, when this formalism is applied to liquid crystal elastomers. Section~\ref{SmADynamics} discusses the dynamics of Sm$A$ elastomers with strain and director.     Sections~\ref{biaxialHydrodynamics} and  ~\ref{SmChydrodynamics} treat, respectively, the hydrodynamics of biaxial smectic and Sm$C$ elastomers. Section~\ref{SmCDynamicsStrainDir} presents our theory for Sm$C$'s with strain and director. Section~\ref{concludingRemarks} contains concluding remarks. There are two appendixes that, respectively, compile results for elastic constants and provide some details of our calculations.

\section{Dynamical equations for liquid crystal elastomers}
\label{dynamicalEquationsReview}
In Ref.~\cite{stenull_lubensky_2004}, where we studied the dynamics of nematic elastomers, we applied the Poisson bracket formalism for coarse-grained variables to derive hydrodynamical equations for the displacement and momentum density as well as equations of motion that in addition involve the Frank director. The validity of these equations is, however, not limited to nematics. 
Our hydrodynamical equations apply for conventional elastomers as well as for the usual (one component) liquid crystal elastomers in, e.g., their isotropic, nematic and smectic phases. Our dynamical equations involving $\brm{n}$ are valid for any liquid crystal elastomer with a defined director, like nematic and smectic elastomers. In this section we review the dynamical equations derived in Ref.~\cite{stenull_lubensky_2004} to provide important background information, to stage known results that we will need as input as we move along and to establish notation.

\subsection{Poisson bracket formalism for coarse-grained variables}
Stochastic dynamical equations for
coarse-grained fields~\cite{HohenbergHal1977,tomsBook} can
be obtained by combining the Poisson-Bracket formalisms of
Classical mechanics \cite{goldstein} and the Langevin \cite{Langevin}
approach to stochastic dynamics. The Poisson-Bracket formalism ensures that the resulting equations of motion are, in the absence of dissipation, invariant under time reversal. The Langevin approach provides a description of dissipative processes and noise forces.  Let
$\Phi_\mu ( \brm{x} ,t )$, $\mu = 1, 2, ...$ be a set of
coarse-grained fields that describe the long-wavelength, low-frequency dynamics of a system, i.e., they are hydrodynamic or quasihydrodynamic variables whose characteristic decay times in the long-wavelength limit are much larger than microscopic decay times. The dynamical equations for $\Phi_\mu ( \brm{x},t)$ are
first-order differential equations in time:
\begin{align}
\label{genHarvard} &\dot{\Phi}_\mu (\brm{x}, t)
\nonumber \\
&= - \int d^d x^\prime \int d t^\prime \, \{ \Phi_\mu
(\brm{x}, t), \Phi_\nu (\brm{x}^\prime, t^\prime) \} \,
\frac{\delta \mathcal{H}}{\delta \Phi_\nu (\brm{x}^\prime,
t^\prime)}
\nonumber \\
& - \Gamma_{\mu, \nu}\, \frac{\delta \mathcal{H}}{\delta
\Phi_\nu (\brm{x}, t)}  \, ,
\end{align}
where $\mathcal{H}$ is the a coarse-grained Hamiltonian that describes the statistical mechanics of the fields $\Phi_\mu ( \brm{x} ,t )$. In Eq.~(\ref{genHarvard}) and in the remainder of this paper the Einstein summation convention is understood. The first term on the right hand side is a
non-dissipative velocity, also known as the reactive term,
that contains the Poisson bracket  $\{ \Phi_\mu (\brm{x},
t), \Phi_\nu (\brm{x}^\prime, t^\prime) \}$ of the coarse
grained fields. The reactive term couples
$\dot{\Phi}_\mu$ to $\delta \mathcal{H}/\delta \Phi_\nu$
only if $\Phi_\mu$ and $\Phi_\nu$ have opposite signs under
time reversal (when external magnetic fields are zero). The
second term on the right hand side is a dissipative term.
$\Gamma_{\mu, \nu}$ is the so-called
dissipative tensor. It may depend on the fields $\Phi_\mu ( \brm{x} ,t )$ and it may contain $- \partial_i \partial_j$, where $\partial_i \equiv \partial / \partial x_i$. Its specifics are determined by three principles:   (i) $\Gamma_{\mu, \nu}$ couples $\dot{\Phi}_\mu$ to $\delta \mathcal{H}/\delta \Phi_\nu$ only if $\Phi_\mu$ and $\Phi_\nu$ have the same sign under time reversal. (ii) By virtue of the Onsager principle~\cite{deGroot}, it must be symmetric in the absence of external magnetic fields. (iii) $\Gamma_{\mu, \nu}$ has to be compatible with the symmetries of the dynamic system. 

When Eq.~(\ref{genHarvard}) is augmented by a noise term $\zeta_\mu$, it represents a stochastic or Langevin equation that cound be used, for example,  to set up a dynamic
functional~\cite{Ja76,DeDo76,Ja92} to study the effects of nonlinearities and fluctuations via dynamical field theory. In this paper, we will ignore the effects of noise.

\subsection{Hydrodynamic equations}
Pure hydrodynamics describes the leading low-frequency, long-wavelength behavior of a system, i.e., it accounts exclusively for those degrees of freedom whose characteristic frequencies $\omega$ vanish in the limit of vanishing wavenumber. There are two general classes of hydrodynamic variables:
conserved variables and broken symmetry variables. A single-component liquid crystal elastomer has five conserved variables, viz.\ the energy density $\epsilon$, the mass density $\rho$, and the three components of the momentum density $\brm{g}$. There are three broken-symmetry variables, namely the three components of the elastic displacement. Throughout this paper, we will use Lagrangian coordinates in which $\brm{x}$ labels a mass point in the unstretched (reference) material and $\brm{R}(\brm{x})$ labels the position of the mass point $\brm{x}$ in the stretched (target) material, so that the broken-symmetry displacement variable is $\brm{u}(\brm{x}) = \brm{R}(\brm{x})- \brm{x}$. Since liquid crystal elastomers are, as any elastomers, permanently crosslinked, the mass density and the displacement are not independent. Rather, changes $\delta \rho$ in mass density are locked to changes in volume such that in the linearized limit $\delta \rho/\rho = - {\bm \nabla} \cdot \brm{u}$. Thus, there are in total $7$ independent hydrodynamic variables and $7$ associated hydrodynamic modes. 

In the following, we will consider only isothermal processes so that the energy density, which is associated with heat diffusion, can be ignored. This
leaves us with six hydrodynamical variables, the components of the momentum
density $g_i(\brm{x})$ and the displacement $u_i(\brm{x})$.
These are independent variables at the reference point
$\brm{x}$ that satisfy the continuum generalizations of the
usual relations for the momentum and displacement of a
particle: $\delta g_i (\brm{x})/\delta g_j (\brm{x}') =
\delta_{ij} \delta ( \brm{x} - \brm{x}')$, $\delta u_i
(\brm{x})/\delta u_j (\brm{x}') = \delta_{ij} \delta (
\brm{x} - \brm{x}')$, and $\delta g_i (\brm{x})/\delta
u_j (\brm{x}') = 0$.  These relations yield a non-vanishing
Poisson bracket between $u_i ( \brm{x})$ and
$g_j(\brm{x}')$:  $\{u_i ( \brm{x}),g_j(\brm{x}')\} = \delta_{ij}
\delta(\brm{x} - \brm{x}')$.  The $\brm{g}-\brm{g}$ and $\brm{u}-\brm{u}$ Poisson
brackets are zero. Using these results and the fact that  the coarse-grained kinetic energy reads $\mathcal{H}_{\rm kin} = \int d^3 x \, \brm{g}^2/(2 \rho)$, one obtains the coarse-grained
equations of motion
\begin{subequations}
\label{genStructSimple}
\begin{align}
\label{EOMsimple1a} \dot{u}_i & = \frac{1}{\rho} \, g_i
\, ,
\\
\label{EOMsimple1b}
 \dot{g}_i &=   - \frac{\delta \mathcal{H}}{\delta u_i} +
\eta_{ijkl}\, \partial_j \partial_l \, \dot{u}_k\,  .
\end{align}
\end{subequations}
Here an in the following we use the convention that indices from the middle of the alphabet,
$\{i,j,k,l\}$, run from 1 to 3 (corresponding to the $x$, $y$, and $z$ directions). $\mathcal{H} = \int d^3 x \, f$ is the elastic energy describing the elastomer with $f$ the corresponding elastic energy density. The specifics of $f$ depend on the particular elastomeric phase under consideration and will be discussed as we go along. The absence of any dissipative term proportional to $-\delta \mathcal{H}/\delta u_i$ in Eq.~(\ref{EOMsimple1a}), which would describe permeation, reflects the tethered or crosslinked character of the elastomer. Pure couplings of $\dot{g}_i$ to $\dot{u}_k$ are forbidden because Eqs.~(\ref{genStructSimple}) have to obey Galileian invariance, and thus the components of the dissipative tensor coupling $\dot{g}_i$ to $\dot{u}_k$ are of the form $-\eta_{ijkl}\, \partial_j \partial_l$, with $\eta_{ijkl}$ being the viscosity tensor. In our discussion of the hydrodynamics of the biaxial smectic and the Sm$C$ phase, to be presented in sections~\ref{biaxialHydrodynamics} and \ref{SmChydrodynamics}, we will restrict ourselves for simplicity to frequencies $\omega$ that are much less than the  inverse characteristic time $\tau_E^{-1}$ associated with the viscosities. In this limit, the $\eta_{ijkl}$ can be viewed as local in time, or equivalently, their temporal Fourier transforms can be considered as being constant.    

To discuss the dynamics of the individual smectic phases in detail, it will be convenient to combine Eqs.~(\ref{EOMsimple1a}) and (\ref{EOMsimple1b}) into a set of second order differential equation for the displacement components $u_i$ only. Switching from time to frequency space via Fourier transformation, this set of equations can be expressed as 
\begin{align}
\label{compactEOMs}
\rho \omega^2 u_i = - \partial_j \sigma_{ij} (\omega)  \, ,
\end{align}
where $\sigma_{ij} (\omega)$ are the components of the stress tensor $\tens{\sigma}$. The specifics of the stress tensor, which  depend on $f$ and the viscosity tensor, will be stated for the individual phases as we proceed. 

\subsection{Dynamic equations with displacement and director}
When the Frank director is included as a dynamical variable, there is an additional non-vanishing Poisson bracket, viz.\ $\{n_i(\brm{x}), g_j (\brm{x}^\prime)\}= -\lambda_{ijk}
\partial_k \delta (\brm{x} - \brm{x}')$~\cite{Forster1983,stark_lubensky_03}. Then, Eq.~(\ref{genHarvard}) leads to the equations of motion
\begin{subequations}
\label{genStruct}
\begin{eqnarray}
\label{EOMa} \dot{n}_i &=& \lambda_{ijk} \, \partial_k
\dot{u}_j - \Gamma \, \frac{\delta \mathcal{H}}{\delta n_i}
\, ,
\\
\label{EOMb} \dot{u}_i &=& \frac{1}{\rho} \, g_i \, ,
\\
\label{EOMc} \dot{g}_i &=& \lambda_{jik} \, \partial_k
\frac{\delta \mathcal{H}}{\delta n_j}  - \frac{\delta
\mathcal{H}}{\delta u_i} + \nu_{ijkl}\, \partial_j
\partial_l \dot{u}_k \, .
\end{eqnarray}
\end{subequations}
The properties of the tensor $\lambda_{ijk}$ are dictated
by three constraints: First, the magnitude of the director
has to be conserved, i.e., $\brm{n} \cdot \dot{\brm{n}} =0$
implying  $n_i \lambda_{ijk} = 0$; second, the equations of
motion must be invariant under $\brm{n} \rightarrow -
\brm{n}$ implying $\lambda_{ijk}$ must change sign with
$\brm{n}$; and third, under rigid uniform rotations, the
director has to obey $\dot{\brm{n}} = \frac{1}{2} (\nabla
\times \dot{\brm{u}}) \times \brm{n}$. The constrains imply that $\lambda_{ijk}$ is of the form
\begin{eqnarray}
\lambda_{ijk} = \frac{\lambda}{2} \, \left( \delta_{ij}^T
\, n_k  + \delta_{ik}^T \,n_j \right)- \frac{1}{2}
\, \left( \delta_{ij}^T \, n_k  - \delta_{ik}^T \,n_j
\right) ,
\end{eqnarray}
where $\delta_{ij}^T = \delta_{ij} - n_i n_j$ is the projector on the subspace perpendicular to $\brm{n}$. There are no constraints on the value of $\lambda$. The second term on the right hand side of Eq.~(\ref{EOMa}) is a dissipative term that describes diffusive relaxation of the director. Its coefficient $\Gamma$ has the dimensions of an inverse viscosity. As before, there are no dissipative contributions to Eq.~(\ref{EOMb}) because liquid crystal elastomers are tethered. There is no dissipative contribution to Eq.~(\ref{EOMa}) involving $\delta \mathcal{H}/\delta u_i$ because the terms on the right side of the equation must be odd in $n_i$ owing to $\brm{n} \rightarrow -\brm{n}$ symmetry. Any such terms must be of the form $A_{ij}\, \delta \mathcal{H}/\delta u_j$, and it is impossible to construct a tensor $A_{ij}$ that satisfying the symmetries of the system and making $A_{ij}\, \delta \mathcal{H}/\delta u_j$ odd under $\brm{n} \rightarrow -\brm{n}$.  A similar argument rules out a dissipative contribution to Eq.~(\ref{EOMb}) that involves $\delta \mathcal{H}/\delta n_j$. As before, Galileian invariance demands that the dissipative coupling of $\dot{g}_i$ to $\dot{u}_k$ are of the form $-\nu_{ijkl}\, \partial_j \partial_l$. Here we use the notation $\nu_{ijkl}$ instead of $\eta_{ijkl}$ for the viscosity tensor, so that we can cleanly keep
track of differences between purely hydrodynamical theories and theories with strain and director. In general, we will assume that $\Gamma$ and the $\nu_{ijkl}$ are, like the $\eta_{ijkl}$, independent of frequency. In Sec.~\ref{SmCrheology}, where we discuss the behavior of Sm$C$ elastomers in rheology experiments, however, we will refrain from this simplifying assumption and we will replace the constants $\Gamma$ and the $\nu_{ijkl}$ by phenomenological functions $\Gamma (\omega)$ and $\nu_{ijkl} (\omega)$ of frequency. 

\section{Smectic-$A$ elastomers -- Dynamics with strain and director}
\label{SmADynamics}
Macroscopically, Sm$A$ elastomers have uniaxial symmetry. When they are crosslinked, as we assume, in the Sm$A$ phase, the uniaxial symmetry is not spontaneous but rather permanently imprinted and the material cannot possess soft elasticity. In the small deformation limit, these elastomers are macroscopically simply uniaxial rubbers (albeit with a very large value of the modulus $C_1$ for extension or compression along the layer normal). As such, their hydrodynamics is identical to that of uniaxial solids~\cite{LLhydrodynamics} Beyond the hydrodynamic regime, however, one must expect an influence of the director on the dynamics of Sm$A$ elastomers in which the genuine difference between Sm$A$ and conventional uniaxial elastomers shows up. Here we apply our dynamical equations with displacement and director to the Sm$A$ phase in order to study these effects.

\subsection{Elastic energy}
As alluded to above, Sm$A$ elastomers are macroscopically simply uniaxial rubbers, at least when strains are small. Therefore, the harmonic stretching elastic energy density $f_{\brm{u}}$ of a Sm$A$ elastomer has exactly the same form as that of a uniaxial solid~\cite{ashcroft_mermin}:
\begin{align}
\label{uniax}
f_{\brm{u}}& = 
 \textstyle{\frac{1}{2}}\, C_1 \, u^2_{zz} + C_2 \, u_{zz} u_{ii} +
 \textstyle{\frac{1}{2}}\,C_3 \, u_{ii}^2
\nonumber \\
& + C_4 \, u_{ab}^2 + C_5 \, u_{az}^2  \, .
\end{align}
We omit higher order terms in the strain because they are inconsequential for our linearized dynamical equations. We have chosen the coordinate system so that the $z$-axis lies in the uniaxial direction. Here and in the following, indices from the beginning of the alphabet, $\{a,b\}$, take on the values 1 and 2. The $u_{ij}$ are components of the Cauchy-Saint-Venant \cite{Love1944,Landau-elas} strain tensor $\tens{u}$,
\begin{subequations}
\label{defStrain}
\begin{align}
\label{defStraina}
u_{ij} ( \xv ) &= \textstyle{\frac{1}{2}}\,
(\Lambda^T_{ik}\Lambda_{kj} - \delta_{ij})
 \\
 \label{defStrainb}
&= \textstyle{\frac{1}{2}} \, (\partial_i u_j +\partial_j u_i ) + \cdots  \, ,
\end{align}
\end{subequations}
where $\Lambda_{ij} =\partial R_i /\partial x_j$ are components of the Cauchy deformation tensor $\tens{\Lambda}$. In Eq.~(\ref{defStrainb}) we have dropped the nonlinear part of the strain tensor, because, for our purposes, it is sufficient to work to harmonic order. 

For setting up a dynamical theory for Sm$A$ elastomers that goes beyond the hydrodynamic limit, we need a more detailed level of description that explicitly involves the Frank director $\brm{n}$. When $\brm{n}$ is included, the overall elastic energy density $f$ will have the following contributions:
\begin{eqnarray}
\label{fullEnergy} 
f = f_{\brm{u}} + f_{\brm{n}-\text{tilt}} + f_{\brm{n}-\text{Frank}} +f_{\text{coupl}} \, ,
\end{eqnarray}
with $f_{\brm{u}}$ as given in Eq.~(\ref{uniax}). $f_{\brm{n}-\text{tilt}}$ is a tilt energy density that accounts for the energy cost when the director, which in an equilibrium Sm$A$ elastomer is aligned along the normal $\brm{N}$ of the smectic layers, tilts away from $\brm{N}$. $f_{\brm{n}-\text{Frank}}$ is the density of the usual Frank energy that describes deviations from the homogeneous equilibrium orientation of the mesogenic component. $f_{\text{coupl}}$ finally is the energy density of coupling between strain and director distortions. 

Coupling the strain and the director brings about an intricate conceptual problem: whereas the liquid crystalline fields  $\brm{N}$ and $\brm{n}$ transform as (rank 1) tensors in target space and are scalars with respect to rotations in reference space, the strain tensor $\tens{u}$ is a (rank 2) tensor in reference space, and it is a scalar in target space. Thus, in order to construct
meaningful combinations of $\tens{u}$ and the liquid crystalline
fields, we have to de able to represent these quantities in either
space. The matrix polar decomposition theorem~\cite{HornJoh1991}
provides a route to this representation. It allows us convert (or
rotate) any reference space vector $\tilde{\brm{b}}$ to a target
space vector $\brm{b}$ via 
\begin{eqnarray}
\label{referenceToTarget}
\brm{b}=\tens{O} \cdot \tilde{\brm{b}}
\end{eqnarray}
 and a target-space vector to a reference space
vector via 
\begin{eqnarray}
\label{targetToRefernce}
\tilde{\brm{b}}= \tens{O}^T\cdot \brm{b} \, , 
\end{eqnarray}
with the ration matrix $\tens{O}$ given by
\begin{eqnarray}
\label{conversionMatrix}
\tens{O} = \tens{\Lambda}\, \big( \tens{\Lambda}^T \tens{\Lambda} \big)^{-1/2} = \tens{1} + \tens{\eta}_A  + \cdots  \, ,
\end{eqnarray}
where $\tens{1}$ is the unit matrix and where $ \tens{\eta}_A = \frac{1}{2} (\tens{\eta} - \tens{\eta}^T)$ is the antisymmetric part of the displacement gradient tensor $\tens{\eta}$ defined by $\eta_{ij} = \partial_j u_i$. In particular, we can rotate the target space director $\nv\equiv (\brm{c}, n_z)\, ,
\quad n_z = \sqrt{1 - c_a^2}$, where $\brm{c}$ is the so-called
c-director, to a reference space vector $\tilde{\nv}\equiv
(\tilde{\brm{c}}, \tilde{n}_z)\, , \quad \tilde{n}_z = \sqrt{1 -
\tilde{c}_a^2}$, and rotate the unit vector $\tilde{\brm{e}}=
(0,0,1)$ specifying the direction of the uniaxial axis in the
reference space to a target-space unit vector.

With the use of the polar decomposition technique, it is not difficult to formulate the tilt and the coupling energy in reference space variables. As mentioned above, in the
equilibrium Sm$A$ phase, the director prefers to be parallel to
both the layer normal $\Nv$ and the anisotropy axis $\ev$~\cite{footnoteAWn0}, which are parallel to each other. There are energy costs proportional to $(\tilde{\Nv}\cdot \tilde{\nv})^2$ and
$(\tilde{\ev}\cdot \tilde{\nv})^2$ associated with deviations from
this equilibrium. These combine to yield a contribution to the tilt energy proportional to $\tilde{c}_a^2$ and higher order terms that are unimportant here. Therefore, the leading contribution to the tilt energy density is
\begin{align}
f_{\brm{n}-\text{tilt}} =\textstyle{\frac{1}{2}} \, r  \, \tilde{c}_a^2 =\textstyle{\frac{1}{2}} \, r  \, Q_a^2 + \cdots \, ,
\end{align}
where $Q_a = c_a - \eta_{Aaz}$. At harmonic order, there is only one contribution to the coupling energy that is compatible with the uniaxial symmetry, viz.\ 
\begin{align}
f_{\text{coupl}} =  \lambda_4 \, \tilde{c}_a u_{az} =   \lambda_4 \, Q_a u_{az} + \cdots \, .
\end{align}
(the symbols $\lambda_1$, $\lambda_2$ and $\lambda_3$ are usually reserved for third-order couplings that, however, do not matter here). When $f_{\brm{n}-\text{tilt}}$ and $f_{\text{coupl}}$ are combined, the result is identical in form to the harmonic coupling energy density
\begin{align}
\label{HunSimple}f_{\brm{u},\brm{n}}& = \textstyle{\frac{1}{2}} \, D_1 \,  Q_a^2  + D_2  \, u_{za} Q_a \, ,
\end{align}
appearing in the elastic energy density
\begin{eqnarray}
\label{nematicEnergy}
f_\text{nem} = f_{\brm{u}} + f_{\brm{n}-\text{Frank}} + f_{\brm{u},\brm{n}} \, ,
\end{eqnarray}
for a nematic elastomer, when the latter is described in terms of strains and the Frank director~\cite{de_Gennes-1}. The stretching energy density $f_{\brm{u}}$ and the Frank energy density $f_{\brm{n}-\text{Frank}}$ in Eq.~(\ref{nematicEnergy}) are the same as in Eq.~(\ref{fullEnergy}), albeit with the elastic constant having, in general, different values. Moreover, $r$ and $\lambda_4$ will be different from $D_1$ and $D_2$. In nematic elastomers, the values of $D_1$ and $D_2$ are such that the shear modulus $C_5$ is renormalized by director relaxation to a value $C_{5-\text{nem}}^R = C_5 - D_2^2/D_1$ that is zero for an ideally soft nematic elastomer or nearly so for a non-ideal semi-soft nematic elastomer. The corresponding renormalized shear modulus $C_{5-\text{Sm$A$}}^R = C_5 - \lambda_4^2/r$ of a Sm$A$ elastomer, to the contrary, will always be significantly larger than zero, when the elastomer was crosslinked, as we assume, in the Sm$A$ phase.

\subsection{Low-frequency, long-wavelength dynamics}
Inserting the just discussed elastic energy of Sm$A$ elastomers into the equations of motion with displacement and director, Eq.~(\ref{genStruct}), we obtain dynamical equations for Sm$A$'s that are identical in form to those of nematic elastomers. The only differences reside in the magnitude of the elastic constants, viscosities and nonhydrodynamic relaxation times. In addition, $C_{5-\text{nem}}^R$ can vanish in soft nematics but $C_{5-\text{Sm$A$}}^R$ is always significantly larger than zero in Sm$A$ elastomers crosslinked in the Sm$A$ phase. Thus, solving the Sm$A$ equations leads to sound velocities, modes etc.\ that have exactly the same from as in nematics. To avoid undue repetition, we will refrain from reviewing these results here in detail; rather we refer the reader directly to Ref.~\cite{stenull_lubensky_2004}. 

\section{Biaxial smectic elastomers -- Hydrodynamics}
\label{biaxialHydrodynamics}
When the shear modulus $C_4$ in the elastic energy density~(\ref{uniax}) becomes negative, as it will in response of biaxial ordering of the constituent mesogens of a Sm$A$ elastomer, the system undergoes a phase transition to a broken-symmetry state with $D_{2h}$ (orthorhombic) symmetry~\cite{stenull_lubensky_SmCsoftness_2005, stenull_lubensky_SmCsoftness_2006}. This mechanism, at least in principle, could produce a soft or semi-soft biaxial smectic elastomer~\cite{stenull_lubensky_SmCsoftness_2005, stenull_lubensky_SmCsoftness_2006,adams_warner_SmCsoftness_2005}. In this section we study the  dynamics of these elastomers. In the framework of Lagrange elasticity theory, a model for their elastic energy density exists thus far in a strain-only formulation but not in a formulation with strain and director. Thus, we will restrict ourselves here to pure hydrodynamics.

\subsection{Elastic energy, viscosity tensor and hydrodynamical equations}
A biaxial smectic elastomer is macroscopically an elastic body with $D_{2h}$ symmetry. Therefore, its stretching energy is that of orthorhombic systems~\cite{ashcroft_mermin}, albeit with an important difference: the modulus for shears in the plane of the smectic layers (with our choice of coordinates the $xy$-plane), $C_{xyxy}^R$, vanishes for an ideally soft biaxial smectic elastomer or is small for a semi-soft one. For details on the derivation of the stretching energy density of soft biaxial smectic elastomers, we refer to Refs.~\cite{stenull_lubensky_SmCsoftness_2005, stenull_lubensky_SmCsoftness_2006}. If $C_{xyxy}^R$ vanishes, shears $u_{xy}$ in the $xy$-plane cost no elastic energy and, therefore, cause no restoring forces. Thus, bending terms proportional to $(\partial_y^2  u_x)^2$ and $(\partial_x^2  u_y)^2$ have to be added to the harmonic elastic energy to ensure mechanical stability. This leads to a overall elastic energy density of the form
\begin{align}
\label{biaxEn}
f &= \textstyle{\frac{1}{2}} \, C_{zzzz}\,
u_{zz}^2 +  \textstyle{\frac{1}{2}} \, C_{xzxz} \,
u_{xz}^2 +  \textstyle{\frac{1}{2}} \, C_{yzyz} \,
u_{yz}^2
\nonumber \\
&+ C_{zzxx} \, u_{zz} u_{xx} + C_{zzyy} \,
u_{zz} u_{yy} + \textstyle{\frac{1}{2}} \,
C_{xxxx}\, u_{xx}^2
\nonumber \\
&+\textstyle{\frac{1}{2}} \, C_{yyyy}\, u_{yy}^2 +
C_{xxyy} \, u_{xx} u_{yy} + \textstyle{\frac{1}{2}} \, C_{xyxy}^R u_{xy}^2
\nonumber \\
&+ \textstyle{\frac{1}{2}} \, B_1\, \big( \partial_y^2  u_x \big)^2 +  \textstyle{\frac{1}{2}} \, B_2\, \big( \partial_x^2  u_y \big)^2  \, ,
\end{align}
with bending moduli $B_1$ and $B_2$. 

Like the elastic constant tensor $C_{ijkl}$, the viscosity tensor $\eta_{ijkl}$ of an orthorhombic system has 9 independent parameters. It can be parameterized so that the entropy production density $T \dot{s} $ takes on the same form as the stretching energy density, with $C_{ijkl}$ replaced by $\eta_{ijkl}$ and $u_{ij}$ replaced by the linearized form of $\dot{u}_{ij}$, i.e.\ $\dot{u}_{ij} =  \partial_i \dot{u}_{j} +  \partial_j \dot{u}_{i}$, 
\begin{align}
\label{biaxEntropy}
T \dot{s}  &= \textstyle{\frac{1}{2}} \, \eta_{zzzz}\,
\dot{u}_{zz}^2 +  \textstyle{\frac{1}{2}} \, \eta_{xzxz} \,
\dot{u}_{xz}^2 +  \textstyle{\frac{1}{2}} \, \eta_{yzyz} \,
\dot{u}_{yz}^2
\nonumber \\
&+\eta_{zzxx} \, \dot{u}_{zz} \dot{u}_{xx} + \eta_{zzyy} \,
\dot{u}_{zz} \dot{u}_{yy} + \textstyle{\frac{1}{2}} \,
\eta_{xxxx}\, \dot{u}_{xx}^2
\nonumber \\
&+\textstyle{\frac{1}{2}} \, \eta_{yyyy}\, \dot{u}_{yy}^2 +
\eta_{xxyy} \, \dot{u}_{xx} \dot{u}_{yy} + \textstyle{\frac{1}{2}} \, \eta_{xyxy}^R \dot{u}_{xy}^2  \, .
\end{align}
Equation~(\ref{biaxEntropy}) contains all contributions to $T \dot{s}$ in the hydrodynamic limit. In writing it, we have assumed that all non-hydrodynamic degrees of freedom, like the director, have relaxed to their local equilibrium values in the presence of $\dot{u}_{ij}$. $\eta_{xyxy}^R$ is an effective shear viscosity that is, like the corresponding shear modulus $C_{xyxy}^R$, renormalized by the relaxation of the director.

Now we possess all the ingredients that are required to write down the hydrodynamical equations for biaxial smectics. Inserting the elastic energy~(\ref{biaxEn}) into Eq.~(\ref{genStructSimple}) and using the form of the viscosity tensor implied in Eq.~(\ref{biaxEntropy}), we obtain the equations of motion in the form of Eq.~(\ref{compactEOMs}) with the components of the stress tensor $\tens{\sigma}$ given by
\begin{subequations}
\label{biaxialStressTens}
\begin{align}
&\sigma_{xx} (\omega) = C_{xxxx} (\omega) \, u_{xx} + C_{xxyy} (\omega) \, u_{yy}+ C_{xxzz} (\omega)\, u_{zz} \, ,
\\
&\sigma_{yy} (\omega) = C_{xxyy} (\omega)  \, u_{xx} + C_{yyyy} (\omega) \, u_{yy} + C_{yyzz} (\omega) \, u_{zz} \, ,
\\
&\sigma_{zz} (\omega) = C_{xxzz} (\omega)  \, u_{xx} + C_{yyzz} (\omega) \, u_{yy} + C_{zzzz} (\omega) \, u_{zz} \, ,
\\
&\sigma_{xy} (\omega) =  \textstyle{\frac{1}{2}} \, C_{xyxy}^R (\omega) \, u_{xy} - B_1 \partial_y^3 u_x \, ,
\\
&\sigma_{yx} (\omega) =  \textstyle{\frac{1}{2}} \, C_{xyxy}^R (\omega) \, u_{xy} - B_2 \partial_x^3 u_y \, ,
\\
&\sigma_{xz} (\omega) =  \sigma_{zx} (\omega) = \textstyle{\frac{1}{2}} \, C_{xzxz} (\omega) \, u_{xz} \, ,
\\
&\sigma_{yz} (\omega) =  \sigma_{zy} (\omega) = \textstyle{\frac{1}{2}} \, C_{yzyz} (\omega) \, u_{yz} \, .
\end{align} 
\end{subequations}
Here, $C_{ijkl} (\omega)$ stands for
\begin{align}
\label{defFreqDepMod}
C_{ijkl} (\omega) = C_{ijkl} - i \omega \, \eta_{ijkl} \, ,
\end{align} 
and the renormalized frequency dependent modulus $C_{xyxy}^R (\omega)$ is defined as
\begin{align}
C_{xyxy}^R (\omega) = C_{xyxy}^R - i \omega \, \eta_{xyxy}^R \, .
\end{align}
When the renormalized shear modulus $C_{xyxy}^R$ is nonzero, in which case the bending terms featuring $B_1$ and $B_2$ are unimportant for the hydrodynamic behavior, the stress tensor~(\ref{biaxialStressTens}) is identical to the hydrodynamic stress tensor for an orthorhombic solid and therefore the equations of motion~(\ref{compactEOMs}) are dentical to the hydrodynamic equations for an orthorhombic solid in this case.

\subsection{Sound velocities}
\label{biaxialSoundVel}
To assess the mode structure of biaxial smectic elastomers, we start with an
analysis of propagating sound modes in the dissipationless limit, i.e., in the limit where viscosities $\eta_{ijkl}$ are zero. The sound modes have frequencies 
\begin{align}
\label{soundModeFreq}
\omega (\brm{q}) = C (\vartheta,\varphi) \, q \, , 
\end{align} 
where $q = |\brm{q}|$ and where $\vartheta$ and
$\varphi$ are the azimuthal and polar angles of the wavevector $\brm{q}$ in
spherical coordinates. In calculating the sound velocities, one can neglect the bending terms in the stress tensor~(\ref{biaxialStressTens}), even though these are nondissipative. As we will see further below, the bending terms give rise to modes with frequencies $\omega \sim q^2$ along the symmetry directions, where sound velocities vanish, and these modes mix with dissipative ones to become overdamped diffusive modes with $\omega \sim - i q^2$. Therefore, the bending terms do not contribute to the sound velocities of the propagating modes. For simplicity, we will focus here on the ideally soft case with $C_{xyxy}^R = 0$. We will return to the more general case, where $C_{xyxy}^R$ can be nonzero, in Sec.~\ref{biaxialModeStruct}.

\begin{figure}
\includegraphics[width=8.4cm]{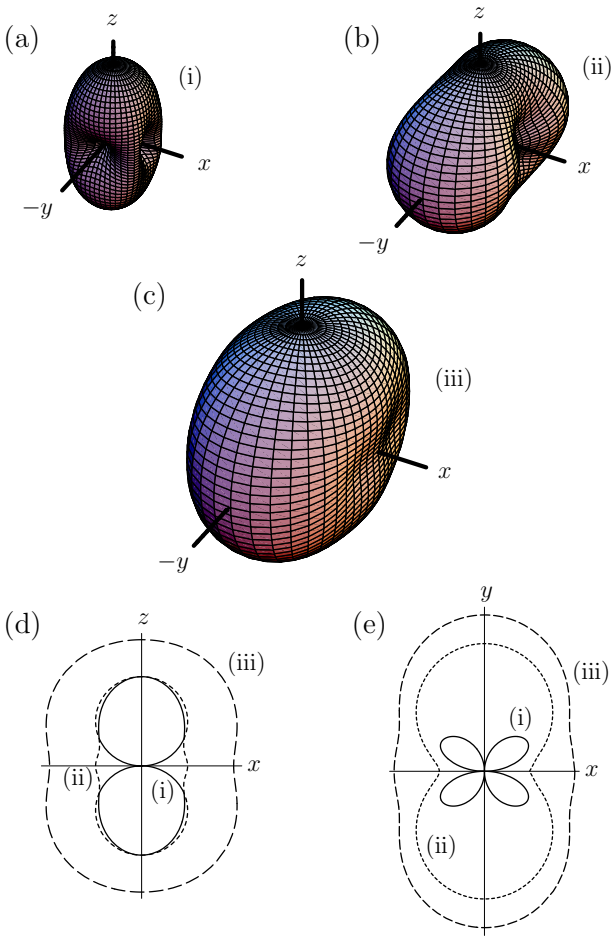}
\caption[]{\label{fig:biaxialSoundVelocities}Schematic plots of sound
velocities. (a), (b) and (c): spherical plots of respectively mode pairs (i), (ii) and (iii). (d) and (e):
polar plots in the $xy$- and $xz$-planes, respectively, of all 3 sound-mode pairs.}
\end{figure}

Setting the viscosities, the bending moduli $K_1$ and $K_2$ and the shear modulus $C_{xyxy}^R$ to zero and switching from the reference space coordinate $\brm{x}$ to the wavevector $\brm{q}$ via Fourier transformation, the hydrodynamical equations simplify to
\begin{subequations}
\label{biaxialDissipationlessEOM}
\begin{align}
\rho \, \omega^2 \, u_x &= \big[ C_{xxxx} \, q_x^2 + \textstyle{\frac{1}{4}} \, C_{xzxz}  \, q_z^2 \big] u_x
+ C_{xxyy} \, q_x q_y \, u_y 
\nonumber \\
&+ \big[ C_{xxzz} + \textstyle{\frac{1}{4}} \, C_{xzxz}   \big] q_x q_z \, u_z \, ,
\\
\rho \, \omega^2 \, u_y &= C_{xxyy} \, q_x q_y \, u_x  + \big[ C_{yyyy} \, q_y^2 + \textstyle{\frac{1}{4}} \, C_{yzyz}  \, q_z^2 \big] u_y 
\nonumber \\
&+ \big[ C_{yyzz} + \textstyle{\frac{1}{4}} \, C_{yzyz}   \big] q_y q_z \, u_z \, ,
\\
\rho \, \omega^2 \, u_z &= \big[ C_{xxzz} + \textstyle{\frac{1}{4}} \, C_{xzxz}   \big] q_x q_z \, u_x  
\nonumber \\
&+ \big[ C_{yyzz} + \textstyle{\frac{1}{4}} \, C_{yzyz} \big] q_y q_z \, u_y 
\nonumber \\
&+ \big[ \textstyle{\frac{1}{4}} \, C_{xzxz}  \, q_x^2 + \textstyle{\frac{1}{4}} \, C_{yzyz}  \, q_y^2 +C_{zzzz} \, q_z^2 \big]  u_z \, .
\end{align} 
\end{subequations}
Inserting the frequency~(\ref{soundModeFreq}) into Eq.~(\ref{biaxialDissipationlessEOM}) and expressing the components of $\brm{q}$ in spherical coordinates, one can solve for the sound velocities $C (\vartheta,\varphi)$. The resulting sound velocities are plotted schematically in Fig.~\ref{fig:biaxialSoundVelocities}. There are 3 pairs of sound modes. One of these pairs (i) is associated with soft shears in the $xy$-plane. Its velocity vanishes for $\brm{q}$ along
the $x$ and $y$ directions, so that when viewed in the $xy$-plane it has a clover-leave shape. The remaining 2 pairs are associated with non-soft deformations. In the incompressible limit, these pairs become purely transverse (ii) and longitudinal (iii), respectively. In the $xy$
and $xz$-planes, their velocities are non-vanishing in all directions.

Figure~\ref{fig:biaxialSoundVelocities} indicates that biaxial smectic elastomers, if they can be produced, have the potential for technological application. Since the mode pair (iii) is purely longitudinal in the incompressible limit, its velocity will be much greater than that of the remaining two pairs. The sound velocities of the remaining pairs, which have different transversal polarizations, are so that the velocity of pair (i) vanishes in directions where the velocity of pair (ii) remains finite. Therefore, a biaxial smectic elastomer could in principle be used to separate the sound mode pair (i) from the sound mode pair (ii), or in other words, such an elastomer could in principle be utilized in making acoustic polarizers. This kind of application has been proposed a while ago for soft nematic elastomers~\cite{terentjev&Co_NEhydrodyn}. In our discussion of the hydrodynamics of soft Sm$C$ elastomers, Sec.~\ref{SmChydrodynamics}, we will see that Sm$C$'s also allow splitting of sound waves. It seems, as if the potential for acoustic polarization is a generic feature of soft liquid crystal elastomers.

\subsection{Mode structure in the incompressible limit}
\label{biaxialModeStruct}
Now we turn to the mode structure, including dissipation, in the incompressible limit. Even in this limit, the equations of motion remain fairly complicated and their solutions have the potential for unpleasant algebraic complexity. Therefore, we will restrict ourselves in the following to the two simplified cases, where either $q_y$ or $q_z$ vanishes, as respectively in Fig.~\ref{fig:biaxialSoundVelocities} (d) and (e). Despite these simplifications, the resulting equations of motion are still to complicated to be solved in closed form. However, since we are interested in the long-wavelength behavior, we can perturbatively determine solutions in the from of a power series in the wavevector. These solutions will turn out to be of the following two types: (i) propagating modes with frequencies
\begin{align}
\label{wp}
\omega_{p} = \pm C\,  q -  i D_p \, q^2
\end{align}
with sound velocities $C$ and diffusion constants~\cite{footnoteDiffConst} $D_p$, and (ii)
diffusive modes with frequencies
\begin{align}
\label{wd}
\omega_{d} = -  i D_d \, q^2 \pm \sqrt{- (D_d \, q^2)^2 + B \,
q^4}
\end{align}
with diffusion constants $D_d$ and bending terms $B$. For $B/D_d^2 \ll 1$ the
diffusive modes split up into slow and fast modes
\begin{subequations}
\label{wsf}
\begin{align}
\omega_{s} &= -  i B/(2D_d) \, q^2 \, ,
\\
\omega_{f} &= -  i 2D_d \, q^2 \, .
\end{align}
\end{subequations}
Specifics of the sound velocities, the diffusion constants and the
bending terms are given in the following.

For $q_y = 0$, the equations of motion reduce to 
\begin{subequations}
\label{biaxialqy=0EOM}
\begin{align}
\rho \, \omega^2 \, u_x &= \big[ C_{xxxx} (\omega) \, q_x^2 + \textstyle{\frac{1}{4}} \, C_{xzxz}  (\omega) \, q_z^2 \big] u_x 
\nonumber \\
&+ \big[ C_{xxzz} (\omega) + \textstyle{\frac{1}{4}} \, C_{xzxz}  (\omega) \big] q_x q_z \, u_z \, ,
\\
\rho \, \omega^2 \, u_y &=  \big[ \textstyle{\frac{1}{4}} \, C_{xyxy}^R (\omega)  \, q_x^2 + \textstyle{\frac{1}{4}} \, C_{yzyz} (\omega) \, q_z^2  + B_2 \, q_x^4\big] u_y  \, ,
\\
\rho \, \omega^2 \, u_z &= \big[ C_{xxzz} (\omega)  + \textstyle{\frac{1}{4}} \, C_{xzxz}  (\omega)   \big] q_x q_z \, u_x  
\nonumber \\
&+ \big[ \textstyle{\frac{1}{4}} \, C_{xzxz} (\omega)   \, q_x^2 +C_{zzzz} (\omega)  \, q_z^2 \big]  u_z \, .
\end{align} 
\end{subequations}
One of the simplifications that we can enjoy for $q_y = 0$ is that the equation of motion for $u_y$ decouples from the equations of motion for $u_x$ and $u_z$, which are coupled to each other. The $u_y$ equation produces a pair of transverse propagating modes polarized in the $y$-direction, $\brm{u} \parallel \tilde{\brm{e}}_y$, with a sound velocity
\begin{align}
\label{biaxialqy=0SoundVel}
C_y &=  \sqrt{\frac{C_{xyxy}^R  \, \hat{q}_x^2 + C_{yzyz}  \, \hat{q}_z^2}{4 \, \rho}} 
\end{align}
and a diffusion constant
\begin{align}
\label{biaxialqy=0DiffConstProp}
D_{p,y} &=  \frac{\eta_{xyxy}^R   \, \hat{q}_x^2 + \eta_{yzyz}  \, \hat{q}_z^2}{8\, \rho}\, ,
\end{align} 
where $\hat{q}_i$ stands for $q_i/q$. For the ideally soft elastomer, $C_{xyxy}^R =0$, the sound velocity vanishes for $q_z=0$, and the above modes become diffusive with a diffusion constant
\begin{align}
\label{biaxialqy=0DiffConst}
D_{d,y} &=  \frac{\eta_{xyxy}^R }{8\, \rho}  \, \hat{q}_x^2
\end{align}
and a bending related constant
\begin{align}
\label{biaxialqy=0BendConst}
B_y &= 64\, \rho\, B_2 \, .
\end{align}
In the limit $64 \, \rho B_2 \ll (\eta_{xyxy}^R)^2$, these diffusive modes for $q_z=0$ split up into a slow and a fast diffusive mode with frequencies
\begin{subequations}
\label{biaxialqy=0fastAndSlow}
\begin{align}
\omega_{s,y} &= -i \, \frac{4 \, B_2}{\eta_{xyxy}^R}\, q_x^2 \, ,
\\
\omega_{f,y} &= -i \, \frac{\eta_{xyxy}^R}{4\, \rho}\, q_x^2  \, .
\end{align} 
\end{subequations}

The equations of motion for $u_x$ and $u_z$ can be solved by decomposing $(u_x,u_z)$ into a
longitudinal part $u_l$ along $\brm{q}$ and a transversal part
$u_T$. In the incompressible limit $u_l$ vanishes. The equation of
motion for $u_T$ produces a pair of propagating transversal modes with polarization $\brm{u} \parallel \tilde{\brm{e}}_T$, where $\tilde{\brm{e}}_T = (\hat{q}_z, 0, -\hat{q}_x)$. This mode pair has the sound velocity
\begin{align}
\label{biaxialqy=0uTSoundVel}
&C_{T} =\sqrt{\frac{4 ( C_{xxxx} +  C_{zzzz} - 2  C_{xxzz})\hat{q}_x^2 \hat{q}_z^2
+  C_{xzxz}(\hat{q}_x^2 - \hat{q}_z^2)^2}{4\, \rho}}
\end{align}
and a diffusion constant
\begin{align}
\label{biaxialqy=0uTDiffConst}
&D_{p,T} =  \frac{\eta_{xxxx} +  \eta_{zzzz} - 2  \eta_{xxzz}}{2\, \rho} \,\hat{q}_x^2 \hat{q}_z^2 +  \frac{\eta_{xzxz} }{8\, \rho} \, (\hat{q}_x^2 - \hat{q}_z^2)^2  .
\end{align}

We complete our discussion of the hydrodynamics of biaxial smectic elastomers by turning to the case that $\brm{q}$ lies in the $xy$-plane. For $q_z= 0$, the equation of motion for $u_z$ decouples form those for $u_x$ and $u_y$, which remain coupled to one another. The $u_z$-equation yields a pair of transverse propagating modes polarized along $\tilde{\brm{e}}_z$ with sound velocities and diffusion constants, which are of the same form as those given in Eqs.~(\ref{biaxialqy=0SoundVel}) and (\ref{biaxialqy=0DiffConstProp}), albeit with $\hat{q}_z$ replaced by $\hat{q}_y$ as well as $C_{xyxy}^R$ and $\eta_{xyxy}^R$ replaced by $C_{xzxz}$ and $\eta_{xzxz}$, respectively:
\begin{subequations}
\begin{align}
C_z &=  \sqrt{\frac{C_{xzxz}  \, \hat{q}_x^2 + C_{yzyz}  \, \hat{q}_y^2}{4 \, \rho}} \, ,
\\
D_{p,z} &=  \frac{\eta_{xzxz}   \, \hat{q}_x^2 + \eta_{yzyz}  \, \hat{q}_y^2}{8\, \rho}\, .
\end{align}
\end{subequations}
Since both elastic constants contributing to the sound $C_z$ are nonzero, even for an ideally soft biaxial smectic elastomer, this sound velocity is nonzero in any direction in the $xy$-plane. 

To solve the coupled equations for $u_x$ and $u_y$, we proceed as above, and we decompose $(u_x,u_z)$ into a longitudinal part $u_l$ and a transversal part $u_T$, which now is along  
$\tilde{\brm{e}}_T = (\hat{q}_y, 0, -\hat{q}_x)$. In the incompressible limit, $u_l$ is suppressed, and of the two coupled mode pairs only the transverse pair polarized along $\tilde{\brm{e}}_T$ survives. The sound velocity and the diffusion constant of this pair are, respectively, identical to Eqs.~(\ref{biaxialqy=0uTSoundVel}) and (\ref{biaxialqy=0uTDiffConst}), however with $\hat{q}_z$ replaced by $\hat{q}_y$, with $C_{zzzz}$, $C_{xxzz}$ and $C_{xzxz}$, respectively, replaced by $C_{yyyy}$, $C_{xxyy}$ and $C_{xyxy}^R$, and with corresponding replacements for the viscosities:
\begin{subequations}
\begin{align}
&C_{T} =\sqrt{\frac{4 ( C_{xxxx} +  C_{yyyy} - 2  C_{xxyy})\hat{q}_x^2 \hat{q}_y^2
+  C_{xyxy}^R(\hat{q}_x^2 - \hat{q}_y^2)^2}{4\, \rho}}
\\
&D_{p,T} =  \frac{\eta_{xxxx} +  \eta_{yyyy} - 2  \eta_{xxyy}}{2\, \rho} \,\hat{q}_x^2 \hat{q}_y^2 +  \frac{\eta_{xyxy}^R }{8\, \rho} \, (\hat{q}_x^2 - \hat{q}_y^2)^2  .
\end{align}
\end{subequations}
Since now $C_{xyxy}^R$ appears in the sound velocity, the sound velocity vanishes for the ideally soft elastomer if $q_x=0$ or $q_y=0$. In either case, the modes become diffusive. For $q_y=0$, the frequencies of these diffusive modes are identical to the frequencies of the diffusive modes discussed further above. Therefore, in the limit $64 \, \rho K_2 \ll (\eta_{xyxy}^R)^2$ one again has a splitting  into slow and fast diffusive modes with frequencies as stated in Eq~(\ref{biaxialqy=0fastAndSlow}). For $q_x=0$, we find
\begin{subequations} 
\begin{align}
\label{biaxialqy=0DiffConstDiff}
D_{d,T} &=  \frac{\eta_{xyxy}^R }{8\, \rho}  \, \hat{q}_z^2 ,
\\
B_T &= 64\, \rho\, B_1 \, .
\end{align}
\end{subequations} 
 In this case, a splitting up into slow and fast diffusive modes occurs for $64 \, \rho B_1 \ll (\eta_{xyxy}^R)^2$. The frequencies of these slow and fast diffusive read
\begin{subequations}
\label{biaxialqz=0fastAndSlow}
\begin{align}
\omega_{s,T} &= -i \, \frac{4 \, B_1}{\eta_{xyxy}^R}\, q_y^2 \, ,
\\
\omega_{f,T} &= -i \, \frac{\eta_{xyxy}^R}{4\, \rho}\, q_y^2  \, .
\end{align} 
\end{subequations}

\section{Smectic-$C$ elastomers -- Hydrodynamics}
\label{SmChydrodynamics}

In this section we investigate the dynamics of Sm$C$ elastomers in the hydrodynamic limit. First we will review their elastic energy density in a model with strain and director~\cite{stenull_lubensky_SmCsoftness_2005, stenull_lubensky_SmCsoftness_2006}. This model will become particularly important in Sec.~\ref{SmCDynamicsStrainDir}. Here, we need a model elastic energy density in terms of strain only, which we derive from the more complete model by integrating out the director. Then we will come to the actual hydrodynamics. We will write down the complete hydrodynamical equations and we will extract the sound velocities and the mode structure.

\subsection{Elastic energy}
\label{SmCelasticEn}
When formulated in terms of strains and the Frank director~\cite{stenull_lubensky_SmCsoftness_2005, stenull_lubensky_SmCsoftness_2006}, the elastic energy density $f$ of a Sm$C$ elastomer crosslinked in the Sm$A$ phase can be divided into three parts
\begin{align}
\label{formOfF}
f = f_{\brm{u}} +  f_{\brm{u}, \brm{n}} +  f_{\brm{n}-\text{Frank}} \, ,
\end{align}
where $f_{\brm{u}}$ is the stretching energy density depending only on $\tens{u}$, $f_{\brm{u},
\brm{n}}$ describes the coupling of the Frank director
$\brm{n}$ to strain variable, and $f_{\brm{n}-\text{Frank}}$ is the density of the Frank energy. In
the following we choose the coordinate system so that the $z$-axis
is parallel to the director of the initial Sm$A$ phase and the
$x$-axis is parallel to the direction of tilt in the resulting equlibrium
Sm$C$ phase. Stated more precisely, we choose our coordinates so that the equilibrium reference space director $\tilde{\brm{n}}^0 = (\tilde{\brm{c}}^0, \tilde{n}_z^0)$ characterizing the undeformed Sm$C$ phase is of the form $\tilde{\brm{n}}^0 = (S, 0, \sqrt{1-S^2})$, with $S = \tilde{c}_x^0$ being the order parameter of the transition. It should be emphasized that, in general, one has to distinguish carefully between the reference-space equilibrium director $\tilde{\brm{n}}^0$ and the physical equilibrium director $\brm{n}^0$. Note from Eqs.~(\ref{referenceToTarget}) and (\ref{conversionMatrix}), however, that to leading order in the displacement gradients $\eta_{ij}$, $\tilde{\brm{n}}^0$ and $\brm{n}^0$ coincide,
\begin{align}
\label{SmCequiDirector}
\brm{n}^0 = (S, 0, \sqrt{1-S^2}) .
\end{align}
For our linearized dynamical theories, only that leading order matters and, therefore, for our current purposes, we can refrain from distinguishing $\tilde{\brm{n}}^0$ and $\brm{n}^0$. 

With our choice of coordinates, $f_{\brm{u}}$ can be written in the same form as the elastic energy density of conventional monoclinic solids~\cite{ashcroft_mermin},
\begin{align}
\label{fu}
f_{\brm{u}} &=  \textstyle{\frac{1}{2}} \, C_{xyxy} \, u_{xy}^2 +
C_{xyzy} \, u_{xy} u_{zy} + \textstyle{\frac{1}{2}} \, C_{zyzy} \,
u_{zy}^2
\nonumber \\
& +\textstyle{\frac{1}{2}} \, C_{zzzz} \, u_{zz}^2  +
\textstyle{\frac{1}{2}} \, C_{xxxx} \, u_{xx}^2+
\textstyle{\frac{1}{2}} \, C_{yyyy} \, u_{yy}^2 
\nonumber \\
&+ \textstyle{\frac{1}{2}} \, C_{xzxz} \, u_{xz}^2
+ C_{zzxx} \, u_{zz} u_{xx} + C_{zzyy} \, u_{zz} u_{yy}    
\nonumber \\
&+ C_{xxyy} \, u_{xx} u_{yy}  + C_{xxxz} \, u_{xx} u_{xz}+ C_{yyxz} \, u_{yy} u_{xz} 
\nonumber \\
&+ C_{zzxz} \, u_{zz} u_{xz} \, ,
\end{align}
but with constraints relating the three elastic constants in the
first row.  These latter constants can be expressed in terms of an
overall elastic constant $\bar{C}$ and an angle $\theta$, which
depends on the order parameter $S$, as $C_{xyxy} = \bar{C} \cos^2
\theta$, $C_{xyzy} = \bar{C} \cos \theta \sin \theta$, and
$C_{zyzy} = \bar{C} \sin^2 \theta$. 

The coupling energy density $f_{\brm{u}, \brm{n}}$ can be stated as
\begin{align}
\label{fun}
&f_{\brm{u}, \brm{n}} =  \textstyle{\frac{1}{2}} \, \Delta \, [
\delta \tilde{c}_y + \alpha  \, u_{xy} + \beta  \, u_{yx} ]^2
\end{align}
where $\Delta$ is a coupling constant, where $\alpha$ and
$\beta$ are dimensionless parameters, and where $\delta \tilde{c}_y$ is the deviation of $\tilde{c}_y$ from its equilibrium value $\tilde{c}_y^0 = 0$. In the following it will be more useful to work with the physical director than with its reference space counterpart. Switching from $\delta \tilde{\brm{n}} = \tilde{\brm{n}} - \tilde{\brm{n}}^0$ to $\delta \brm{n} = \brm{n} - \brm{n}^0$ via the transformation (\ref{referenceToTarget}), we get 
\begin{align}
\label{funRecast}
&f_{\brm{u}, \brm{n}} =  \textstyle{\frac{1}{2}} \, \Delta \, [
Q_y + \alpha  \, u_{xy} + \beta  \, u_{yx} ]^2 \, ,
\end{align}
with the variable $Q_y$ defined by
\begin{align}
\label{defQy}
Q_y =  \delta n_y - S \, \eta_{Ayx} -  \sqrt{1-S^2} \, \eta_{Ayz} \, ,
\end{align}
and where higher order terms in $\eta_{ij}$ have been discarded.

The remaining contribution to the total elastic energy density that we need to discuss here is the Frank energy density. When expanded to harmonic order about the equilibrium director~(\ref{SmCequiDirector}), it becomes
\begin{align}
\label{FrankExpanded}
& f_{\brm{n}-\text{Frank}} 
\nonumber \\
&=  \textstyle{\frac{1}{2}} \, K_{xxxx} \,(\partial_x n_x)^2 +
\textstyle{\frac{1}{2}} \, K_{yyyy} \,(\partial_y n_y)^2 + \textstyle{\frac{1}{2}} \, K_{yxyx} \,(\partial_x n_y)^2
\nonumber \\
& +\textstyle{\frac{1}{2}} \, K_{xyxy} \,(\partial_y n_x)^2 +
\textstyle{\frac{1}{2}} \, K_{xzxz} \,(\partial_z n_x)^2 + \textstyle{\frac{1}{2}} \, K_{yzyz} \,(\partial_z n_y)^2
\nonumber \\
&+ K_{xxyy} \,\partial_x n_x\, \partial_y n_y +  K_{xyyx} \,\partial_x n_y\, \partial_y n_x  
\nonumber \\
&+ K_{xxxz} \,\partial_x n_x\, \partial_z n_x +  K_{yyyx} \,\partial_y n_y\, \partial_y n_x  
\nonumber \\
&+ K_{yxyz} \,\partial_x n_y\, \partial_z n_y +  K_{xyyz} \,\partial_y n_x\, \partial_z n_y \, ,
\end{align}
where we have replaced $\delta \brm{n}$ by $\brm{n}$ for notational simplicity, with the understanding, that it has only two components $n_x$ and $n_y$. We will stick to this abbreviated notation for the remainder of this paper. The elastic constants $K_{xxxx}$ and so on are combinations of the the original Frank constants, describing respectively splay, twist, and bend distortions of the director and the order parameter $S$. For specifics, see appendix~\ref{specificsFrankConst}.

As discussed above, the director is not a genuine hydrodynamic variable. Therefore, it should be integrated out of the elastic energy as long as we focus on pure hydrodynamics. Minimizing $f$ over $n_x$ and $n_y$, we find that these quantities relax in the presence of strain to
\begin{subequations}
\begin{align}
\label{relaxedDirector}
n_x &= 0 \, ,
\\
n_y & = S \, \eta_{Ayx} +  \sqrt{1-S^2} \, \eta_{Ayz} - \alpha \, u_{xy} - \beta \, u_{yz}
\nonumber \\
&+ \frac{1}{2\Delta} \Big[ K_{yxyx} \, \partial_x^2 + K_{yyyy} \, \partial_y^2  + K_{yxyz} \, \partial_z ^2 + 2 K_{yxyx} \, \partial_x \partial_z \Big]
\nonumber \\
& \times \Big\{  S \, \eta_{Ayx} +  \sqrt{1-S^2} \, \eta_{Ayz} - \alpha \, u_{xy} - \beta \, u_{yz} \Big\} \, ,
\end{align}
\end{subequations}
where terms of fourth and higher orders in derivatives have been discarded. Inserting these results into $f$, we obtain an effective elastic energy density of the form
\begin{align}
\label{fEffecticve}
f = f_{\brm{u}} +  f_{\text{bend}} \, ,
\end{align} 
where $f_{\brm{u}}$ remains as given in Eq.~(\ref{fu}), and where $f_{\text{bend}}$ is a bending energy density given by
\begin{align}
\label{BendingEn} 
f_{\text{bend}} &=  \textstyle{\frac{1}{2}} \, B_1 \,(\partial_x^2 u_y)^2 +
\textstyle{\frac{1}{2}} \, B_2 \,(\partial_z^2 u_y)^2 + \textstyle{\frac{1}{2}} \, B_3 \,(\partial_x \partial_z u_y)^2
\nonumber \\
& +B_4 \, u_y \partial_z \partial_x^3  u_y +
B_5  \,u_y \partial_x \partial_z^3  u_y  + \textstyle{\frac{1}{2}} \,B_6 \,(\partial_y^2 u_x)^2
\nonumber \\
&+ \textstyle{\frac{1}{2}} \,B_7 \,(\partial_y^2 u_z)^2 + B_8 \, (\partial_y^2 u_x) (\partial_y^2 u_z) \, .
\end{align}
Specifics of the bending moduli $B_1$, $B_2$ and so on are given in appendix~\ref{specificsBendingMod}.

The elastic energy density~(\ref{fEffecticve}) implies that Sm$C$ elastomers exhibit static soft elasticity. Suppose deformations are not too large so that the bending contribution to Eq.~(\ref{fEffecticve}) may be neglected. Then, due to the above relations among the elastic constants $C_{xyxy}$, $C_{xyzy}$ and $C_{zyzy}$, deformations characterized by Fourier transformed displacements
$\brm{u}\parallel \tilde{\brm{e}}_y$ and wavevectors $\brm{q} \parallel \tilde{\brm{e}}_2 = (-\sin \theta, 0, \cos \theta)$ or, alternatively, $\brm{u} \parallel \tilde{\brm{e}}_2$ and $\brm{q} \parallel \tilde{\brm{e}}_y$ cost no elastic energy and hence cause no restoring forces. In the following, we will occasionally switch to a coordinate system that is rotated relative to the original coordinate system though an angle $\theta$ about the $y$-axis so that axis of the new system are $\tilde{\brm{e}}_{x^\prime} = (\cos \theta, 0, \sin \theta)$, $\tilde{\brm{e}}_{y^\prime} = \tilde{\brm{e}}_y$, and $\tilde{\brm{e}}_{z^\prime} = \tilde{\brm{e}}_2$. One one hand, the effects of soft elasticity are more evident in the rotated system. On the other hand, the directions $\tilde{\brm{e}}_{x^\prime}$ and $\tilde{\brm{e}}_{z^\prime}$ will play a certain role in the mode structure of Sm$C$ elastomers, as we will see below. In the rotated system, the stretching energy density takes on the form
\begin{align}
\label{C2hEnRotated}
f_{\brm{u}} &= \textstyle{\frac{1}{2}} \,
C_{x^\prime y^\prime x^\prime y^\prime} \, (u_{x^\prime
y^\prime})^2 + \textstyle{\frac{1}{2}} \, C_{z^\prime z^\prime
z^\prime z^\prime} \, (u_{z^\prime z^\prime })^2
\nonumber \\
& + \textstyle{\frac{1}{2}} \, C_{x^\prime z^\prime x^\prime
z^\prime} \, (u_{x^\prime z^\prime})^2 + C_{z^\prime z^\prime
x^\prime x^\prime} \, u_{z^\prime z^\prime } u_{x^\prime x^\prime}
\nonumber \\
& + C_{z^\prime z^\prime y^\prime y^\prime } \, u_{z^\prime
z^\prime} u_{y^\prime y^\prime} + \textstyle{\frac{1}{2}} \,
C_{x^\prime x^\prime x^\prime x^\prime} \, (u_{x^\prime x^\prime
})^2
\nonumber \\
& + \textstyle{\frac{1}{2}} \, C_{y^\prime y^\prime y^\prime
y^\prime} \, (u_{y^\prime y^\prime })^2 + C_{x^\prime x^\prime
y^\prime y^\prime} \, u_{x^\prime x^\prime} u_{y^\prime y^\prime}
\nonumber \\
&+ C_{x^\prime x^\prime x^\prime z^\prime} \, u_{x^\prime
x^\prime} u_{x^\prime z^\prime}+ C_{y^\prime y^\prime x^\prime
z^\prime} \, u_{y^\prime y^\prime } u_{x^\prime z^\prime}
\nonumber \\
&+ C_{z^\prime z^\prime x^\prime z^\prime} \, u_{z^\prime
z^\prime} u_{x^\prime z^\prime} \, ,
\end{align}
where $C_{x^\prime y^\prime x^\prime y^\prime}= \bar{C}$, $C_{y^\prime y^\prime
y^\prime y^\prime} = C_{y y y y}$ and where the remaining new elastic
constants are non-vanishing conglomerates of the elastic constants
defined via Eq.~(\ref{fu}) and sines and cosines of $\theta$. Note that, in this coordinate system, the elastic constants $C_{x^\prime y^\prime y^\prime z^\prime}$ and $C_{y^\prime
z^\prime y^\prime z^\prime}$ are zero, and that therefore the elastic energy does not depend at all on $u_{y'z'}$, impyling there is no energy cost for shears in the $z^\prime y^\prime$-plane.  If the elastomer is crosslinked in the Sm$C$ phase, these moduli become nonzero. If they remain small, the elastomer will be semi-soft.

\subsection{Hydrodynamic equations}
\label{SmChydrodynamicEqs}
Knowing the effective elastic energy density of Sm$C$ elastomers in terms of the displacement variable only, Eq.~(\ref{fEffecticve}) in conjunction with Eqs.~(\ref{fu}) and (\ref{BendingEn}), we are almost in the position to write down hydrodynamic equations for Sm$C$ elastomers. The only ingredient that is still missing is the viscosity tensor. We will parametrize this tensor so that the density of the entropy production from viscous stresses takes on a form similar to that of Eq.~(\ref{fu}):
\begin{align}
\label{SmCentropyPrdDens}
T \dot{s} &=  \textstyle{\frac{1}{2}} \, \eta^R_{xyxy} \, \dot{u}_{xy}^2 +
\eta^R_{xyzy} \, \dot{u}_{xy} \dot{u}_{zy} + \textstyle{\frac{1}{2}} \, \eta^R_{zyzy} \,
\dot{u}_{zy}^2
\nonumber \\
& +\textstyle{\frac{1}{2}} \,  \eta_{zzzz} \, \dot{u}_{zz}^2  +
\textstyle{\frac{1}{2}} \,  \eta_{xxxx} \, \dot{u}_{xx}^2+
\textstyle{\frac{1}{2}} \,  \eta_{yyyy} \, \dot{u}_{yy}^2 +
\textstyle{\frac{1}{2}} \,  \eta_{xzxz} \, \dot{u}_{xz}^2
\nonumber \\
&+  \eta_{zzxx} \, \dot{u}_{zz} \dot{u}_{xx} +  \eta_{zzyy} \, \dot{u}_{zz} \dot{u}_{yy}    +
 \eta_{xxyy} \, \dot{u}_{xx} \dot{u}_{yy}
\nonumber \\
& + \eta_{xxxz} \, \dot{u}_{xx} \dot{u}_{xz}+  \eta_{yyxz} \, \dot{u}_{yy} \dot{u}_{xz} +
 \eta_{zzxz} \, \dot{u}_{zz} \dot{u}_{xz} \, .
\end{align}
As in Eq.~(\ref{biaxEntropy}), we assume here that all non-hydrodynamic degrees of freedom including the director have relaxed to their local equilibrium values in the presence of strain, i.e., $ \eta^R_{xyxy}$, $\eta^R_{xyzy}$ and $\eta^R_{zyzy}$ are effective viscosities that have been renormalized by the relaxation of the director.

Collecting, we obtain a set of equations for the displacement components that can be stated in the form of Eq.~(\ref{compactEOMs}) with the components of the stress tensor given by
\begin{subequations}
\label{SmCStressTensStrainOnly}
\begin{align}
\sigma_{xx} (\omega) &= C_{xxxx} (\omega) \, u_{xx} + C_{xxyy} (\omega) \, u_{yy}+ C_{xxzz} (\omega)\, u_{zz} 
\nonumber \\
&+ C_{xxxz} (\omega) \, u_{xz} \, ,
\\
\sigma_{yy} (\omega) &= C_{xxyy} (\omega)  \, u_{xx} + C_{yyyy} (\omega) \, u_{yy} + C_{yyzz} (\omega) \, u_{zz}
\nonumber \\
&+ C_{yyxz} (\omega) \, u_{xz} \, ,
\\
\sigma_{zz} (\omega) &= C_{xxzz} (\omega)  \, u_{xx} + C_{yyzz} (\omega) \, u_{yy} + C_{zzzz} (\omega) \, u_{zz}
\nonumber \\
&+ C_{zzxz} (\omega) \, u_{xz} \, ,
\\
\sigma_{yx} (\omega) &=  \textstyle{\frac{1}{2}} \, C_{xyxy}^R (\omega) \, u_{xy} +  \textstyle{\frac{1}{2}} \, C_{xyyz}^R (\omega) \, u_{yz} - B_1 \, \partial_x^3 u_y
\nonumber \\
& - B_3 \, \partial_x \partial_z^2 u_y - 2\, B_4 \, \partial_x^2 \partial_z u_y - 2\, B_5 \, \partial_z^3 u_y \, ,
\\
\sigma_{yz} (\omega) &=  \textstyle{\frac{1}{2}} \, C_{xyyz}^R (\omega) \, u_{xy} +  \textstyle{\frac{1}{2}} \, C_{yzyz}^R (\omega) \, u_{yz} - B_2 \, \partial_z^3 u_y
\nonumber \\
& - B_3 \, \partial_x^2 \partial_z u_y - 2\, B_4 \, \partial_x^3 u_y - 2\, B_5 \,  \partial_x \partial_z2 u_y \, ,
\\
\sigma_{xy} (\omega) &=  \textstyle{\frac{1}{2}} \, C_{xyxy}^R (\omega) \, u_{xy} +  \textstyle{\frac{1}{2}} \, C_{xyyz}^R (\omega) \, u_{yz} 
\nonumber \\
&- B_6 \, \partial_y^3 u_x - B_8 \, \partial_y^3 u_z \, ,
\\
\sigma_{zy} (\omega) &=   \textstyle{\frac{1}{2}} \, C_{xyyz}^R (\omega) \, u_{xy} +  \textstyle{\frac{1}{2}} \, C_{yzyz}^R (\omega) \, u_{yz} 
\nonumber \\
&- B_7 \, \partial_y^3 u_z - B_8 \, \partial_y^3 u_x \, ,
\\
\sigma_{xz} (\omega) &=  \sigma_{zx} (\omega) = \textstyle{\frac{1}{2}} \, C_{xxxz} (\omega) \, u_{xx} + \textstyle{\frac{1}{2}} \, C_{yyxz} (\omega) \, u_{yy}  
\nonumber \\
&+ \textstyle{\frac{1}{2}} \, C_{zzxz} (\omega)\, u_{zz} + \textstyle{\frac{1}{2}} \, C_{xzxz} (\omega) \, u_{xz} \, ,
\end{align} 
\end{subequations}
where $C_{ijkl} (\omega)$ without superscript $R$ is defined as in Eq.~(\ref{defFreqDepMod}) and
\begin{align}
\label{renormalizedModuliHydrodyn}
C_{\xi \chi}^R (\omega) &= C_{\xi \chi} - i \omega \, \eta_{\xi \chi}^R  \, .
\end{align} 
Here, we have used a compact notation with indices $\xi$, $\chi$ running over the pairs $xy$ and $zy$. We will on occasion return to this notation below.  When the shear moduli $C_{\xi \chi}$ are nonzero and independent, in which case the bending terms  are unimportant for the hydrodynamic behavior, the stress tensor~(\ref{SmCStressTensStrainOnly}) is identical to the hydrodynamic stress tensor for an monoclinic solid.  

\subsection{Sound velocities}
\begin{figure}
\includegraphics[width=8.4cm]{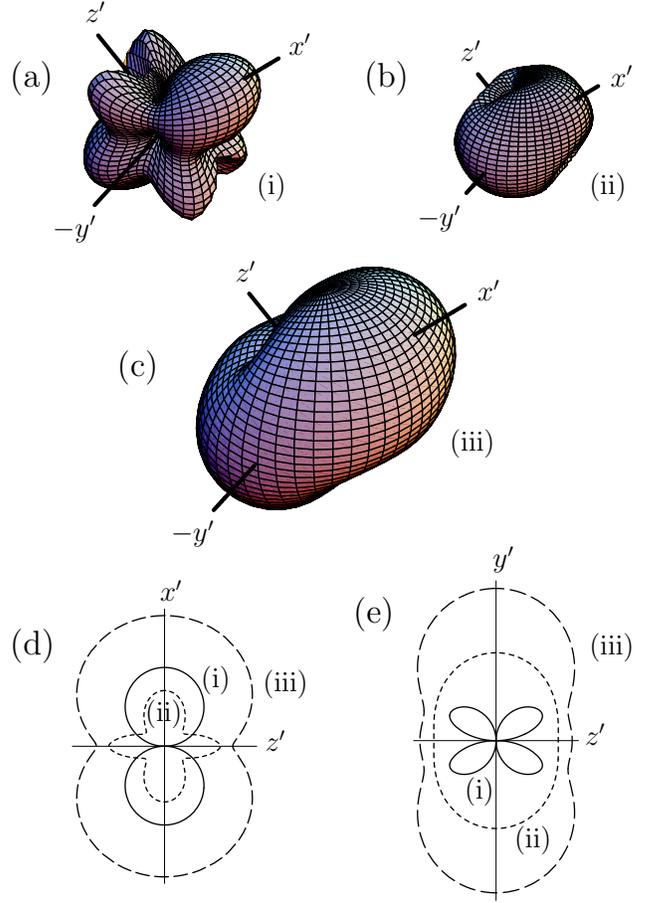}
\caption[]{\label{fig:SmCsoundVelocities}Schematic plots of sound
velocities. (a), (b) and (c): spherical plots of respectively mode pairs (i), (ii) and (iii). (d) and (e):
polar plots in the $x^\prime z^\prime$- and $x^\prime z^\prime$-planes, respectively, of all 3 sound-mode pairs. Plot (a) is magnified relative to the remaining plots by a factor 2.}
\end{figure}

In order to determine the mode structure of Sm$C$ elastomers, we proceed as in our discussion of biaxial smectics by analyzing the sound-mode structure in the dissipationless limit. As before, any bending terms can be neglected in this analysis. Setting all viscosities to zero, we extract from the equations of motion the characteristic frequencies in the form of Eq.~(\ref{soundModeFreq}). The resulting sound velocities are plotted schematically in Fig.~\ref{fig:SmCsoundVelocities}. There are 3 pairs of sound modes. One of these pairs (i) is associated with the soft deformations
discussed in Sec.~\ref{SmCelasticEn}. Its velocity vanishes for $\brm{q}$ along
$\tilde{\brm{e}}_y$ and $\tilde{\brm{e}}_2$ so that when viewed in the $y^\prime
z^\prime$-plane it has a clover leave like shape. The remaining 2
pairs are associated with non-soft deformations. In the
incompressible limit, these pairs become purely transverse (ii)
and longitudinal (iii), respectively. In the $y^\prime z^\prime$
and $x^\prime z^\prime$-planes, their velocities are non-vanishing
in all directions. Note that, since the velocity of pair (i)
vanishes in directions where the velocities of the other modes
remain finite, Sm$C$ elastomers are, like nematic~\cite{terentjev&Co_NEhydrodyn} and biaxial smectic leastomers, potential candidates for applications in acoustic polarizers.

\subsection{Mode structure in the incompressible limit}
\label{SmCModeStructHydrodyn}
Having found the general sound-mode structure in the nondissipative limit, we now turn to the full mode structure in the incompressible limit. Due to the complexity of the equations of motion, we once again focus on the softness-related symmetry directions, cf.\ Fig.~\ref{fig:SmCsoundVelocities} (d) and (e). As in Sec.~\ref{biaxialModeStruct}, the equations of motion yield propagating modes, whose frequencies are of the form of Eq.~(\ref{wp}) and diffusive modes, whose frequencies are of the form of Eqs.~(\ref{wd}) or (\ref{wsf}). Specifics of the sound velocities, the diffusion constants and the
bending terms entering these frequencies are given in the following.

First, let us consider the case that $\brm{q}$ lies in the
$xz$-plane. In this case the equation of motion for $u_y$
decouples from the equations of motion for $u_x$ and $u_z$. The $u_y$
equation produces a pair of transverse propagating modes polarized along $\tilde{\brm{e}}_y$
with a sound velocity
\begin{align}
\label{CyHydro}
C_{y}  &=  \sqrt{\frac{\bar{C}}{4\, \rho}}\,  | \cos \theta \, \hat{q}_x +
\sin \theta \, \hat{q}_z |= \sqrt{\frac{\bar{C}}{4\, \rho}}\, |\hat{q}_{x'}| 
\end{align}
and a diffusion constant
\begin{align}
\label{DpyHydro}
D_{p,y}  &= \frac{\eta_{xyxy}^R \, \hat{q}_x^2 + 2 \, 
\eta_{xyzy}^R\,  \hat{q}_x \hat{q}_z + \eta_{zyzy}^R \, \hat{q}_z^2}{8\, \rho}\, . 
\end{align}
In the soft direction, i.e.\ for $\brm{q} \parallel \tilde{\brm{e}}_{z^\prime}$, the sound velocity $C_y$ vanishes and these propagating modes become diffusive with diffusion constant $D_{d,y} = D_{p,y}$ and bending term
\begin{align}
\label{ByHydro}
B_y  = \frac{B_1 \, \hat{q}_x^4  + B_2 \, 
\hat{q}_z^4 + B_3 \, \hat{q}_x^2 \hat{q}_z^2 + 2 \, B_4\, 
\hat{q}_x^3 \hat{q}_z + 2 \, B_5 \, \hat{q}_x \hat{q}_z^3}{\rho} \, ,
\end{align} 
cf.~Eq.~(\ref{wd}). The equations of motion for $u_x$ and $u_z$ can be solved by decomposing $(u_x,u_z)$ into a longitudinal part $u_l$ along $\brm{q}$ and a transversal part
$u_T$ along $\tilde{\brm{e}}_T = (\hat{q}_z, 0, -\hat{q}_x)$. In the incompressible limit $u_l$ vanishes. The equation of motion for $u_T$ produces a pair of propagating modes with polarization $\brm{u} \parallel \tilde{\brm{e}}_T$, with sound velocity
\begin{align}
\label{CThydro}
C_{T}  &=  \sqrt{1/\rho}\,  \big\{  [C_{xxxx} + C_{zzzz} - 2\, 
C_{xxzz}] \hat{q}_x^2 \hat{q}_z^2
\nonumber  \\
&+ [C_{xxxz}  -  C_{zzxz}] \hat{q}_x \hat{q}_z (\hat{q}_z^2
-\hat{q}_x^2) 
\nonumber \\
&+ \textstyle{\frac{1}{4}}  \, C_{xzxz}(\hat{q}_z^2
-\hat{q}_x^2)\big\}^{1/2}
\end{align}
and diffusion constant
\begin{align}
\label{DpThydro}
D_{p,T} &=  \frac{ \eta_{xxxx} +  \eta_{zzzz} - 2\,  \eta_{xxzz}}{2\, \rho}\,  \hat{q}_x^2 \hat{q}_z^2
\nonumber \\
&+ \frac{ \eta_{xxxz}  -   \eta_{zzxz}}{2\, \rho} \,  \hat{q}_x \hat{q}_z (\hat{q}_z^2
-\hat{q}_x^2) + \frac{ \eta_{xzxz}}{8\, \rho} \, (\hat{q}_z^2-\hat{q}_x^2 )^2 .
\end{align}

Second and last, let us consider the case $\brm{q} \parallel \tilde{\brm{e}}_y$. There
is a pair of longitudinal propagating modes with $\brm{u}
\parallel \tilde{\brm{e}}_y$ that is suppressed in the incompressible
limit. There is a pair of elastically soft diffusive modes polarized along $\tilde{\brm{e}}_{z^\prime}$ with
\begin{subequations}
\begin{align}
\label{DdzPrimeHydro}
& D_{d,z^\prime}  = \frac{\eta_{z^\prime y^\prime z^\prime y^\prime}^R}{8 \, \rho}  \, ,
\\
\label{BzPrimeHydro}
& B_{z^\prime}  = \frac{ \sin^2 \theta \, B_6 -
\sin 2 \theta \,B_8   + \cos^2 \theta \, B_7}{\rho} \, ,
\end{align}
\end{subequations}
where $\eta_{z^\prime y^\prime z^\prime y^\prime}^R = \sin^2 \theta
\,  \eta_{xyxy}^R - \sin 2 \theta \, \eta_{xyzy}^R  + \cos^2 \theta
\, \eta_{zyzy}^R$. Finally, there is a pair of propagating modes polarized
along $\tilde{\brm{e}}_{x^\prime}$ with
\begin{subequations}
\begin{align}
\label{CxPrimeHydro}
& C_{x^\prime}  =  \sqrt{\bar{C}/(4\rho)}\, ,
\\
\label{DpxPrimeHydro}
& D_{p,x^\prime}  =  \frac{ \eta_{x^\prime y^\prime x^\prime y^\prime }^R}{ 8 \, \rho} \,  ,
\end{align}
\end{subequations}
with $\eta_{x^\prime y^\prime x^\prime y^\prime }^R = \cos^2 \theta
\,  \eta_{xyxy}^R + \sin 2 \theta \, \eta_{xyzy}^R  + \sin^2 \theta
\, \eta_{zyzy}^R$.

\section{Smectic-$C$ elastomers -- Dynamics with strain and director}
\label{SmCDynamicsStrainDir}
In this section we will treat the low-frequency, long-wavelength dynamics of Sm$C$ elastomers within our formulation with strain and director.

\subsection{Equations of motion}
Now we will assemble effective equations of motion for the displacement variable $\brm{u}$ and the director. One of the main ingredient entering the general equations of motion~(\ref{genStruct}) is the elastic energy of the Sm$C$ phase. Here we need the elastic energy density with strain and director, Eq.~(\ref{formOfF}), that comprises the stretching energy density $f_{\brm{u}}$, Eq.~(\ref{fu}), the coupling energy density $f_{\brm{u}, \brm{n}}$,  Eq.~(\ref{funRecast}), and the Frank energy density $f_{\brm{n}-\text{Frank}}$, Eq.~(\ref{FrankExpanded}). The other important ingredient is the viscosity tensor. Again, we parametrize this tensor so that the entropy production density is of the form of Eq.~(\ref{SmCentropyPrdDens}), albeit now with $\eta_{ijkl}$ replaced by $\nu_{ijkl}$ (all unrenromalized thus far). 

Recall that we have expanded the coupling and Frank energy densities about the equilibrium director~(\ref{SmCequiDirector}) and that the deviation from the equilibrium director has only two components, which we denote for briefness by $n_x$ and $n_y$. For convenience, we will in the following not work with $n_y$ directly but rather with the composite variable $Q_y$ defined in Eq.~(\ref{defQy}) instead. Then, the two equations of motion resulting from the director equation~(\ref{EOMa}) read
\begin{widetext}
\begin{subequations}
\label{nxQy}
\begin{align}
\label{QyEquation}
[\partial_t + \Gamma \, \Delta + \Gamma \, K_{yy} (\nabla ) ] Q_y + \Gamma \, K_{yx} (\nabla ) \, n_x
&= [\lambda \, S \, \partial_t - \Gamma \, \Delta \, \alpha ] u_{xy}
+  \big[\lambda \, \sqrt{1-S^2} \, \partial_t - \Gamma \, \Delta \, \beta \big] u_{yz} 
\nonumber \\
& -  \Gamma \, K_{yy} (\nabla ) \big\{  S\, \eta_{Ayx} + \sqrt{1-S^2}\, \eta_{Ayz} \big\}  ,
\\
[\partial_t  + \Gamma \, K_{xx} (\nabla ) ] n_x + \Gamma \, K_{xy} (\nabla ) \, Q_y
&= \lambda \, S (1-S^2) \, \partial_t \, [ u_{xx} -u_{zz}] +  \lambda \, (1-2S^2)\sqrt{1-S^2} \, \partial_t\, u_{xz}
\nonumber \\
& - \Gamma \, K_{xy} (\nabla ) \big\{  S\, \eta_{Ayx} + \sqrt{1-S^2}\, \eta_{Ayz} \big\} 
 + \sqrt{1-S^2} \, \partial_t  \eta_{Axz} \, ,
\end{align}
\end{subequations}
where
\begin{subequations}
\label{KxxAndSoOn}
\begin{align}
K_{xx}(\nabla ) &= - K_{xxxx} \, \partial_x^2 - K_{xyxy} \, \partial_y^2 - K_{xzxz} \, \partial_z^2 
- 2\, K_{xxxz} \, \partial_x \partial_z \, ,
\\
K_{yy}(\nabla ) &= - K_{yxyx} \, \partial_x^2 - K_{yyyy} \, \partial_y^2 - K_{yzyz} \, \partial_z^2 
- 2\, K_{yxyz} \, \partial_x \partial_z \, ,
\\
K_{xy}(\nabla ) &= - (K_{xxyy} + K_{xyyx})  \, \partial_x \partial_y 
- (K_{yyxz} + K_{xyyz}) \,  \partial_y \partial_z \, ,
\\
K_{yx}(\nabla ) &= - (K_{xxyy} + K_{xyyx})  \, \partial_x \partial_y 
- (K_{yyxz} + K_{yxyz}) \,  \partial_y \partial_z \, .
\end{align}
\end{subequations}
The equations of motion resulting from the momentum density and displacement equations~(\ref{EOMb}) and (\ref{EOMb}) can be written in the form of Eq.~(\ref{compactEOMs}) with a stress tensor given by
\begin{subequations}
\label{SmCStressTensStrainAndDir}
\begin{align}
\sigma_{xx} (\omega) &= C_{xxxx} (\omega) \, u_{xx} + C_{xxyy} (\omega) \, u_{yy}+ C_{xxzz} (\omega)\, u_{zz}  + C_{xxxz} (\omega) \, u_{xz}  
\nonumber \\
&+ \lambda \, S (1-S^2) K_{xx}(\nabla ) \, n_x + \lambda \, S (1-S^2) K_{xy}(\nabla )  \big[  Q_y + S\, \eta_{Ayx} + \sqrt{1-S^2}\, \eta_{Ayz} \big]  \, ,
\\
\sigma_{yy} (\omega) &= C_{xxyy} (\omega)  \, u_{xx} + C_{yyyy} (\omega) \, u_{yy} + C_{yyzz} (\omega) \, u_{zz} + C_{yyxz} (\omega) \, u_{xz} \, ,
\\
\sigma_{zz} (\omega) &= C_{xxzz} (\omega)  \, u_{xx} + C_{yyzz} (\omega) \, u_{yy} + C_{zzzz} (\omega) \, u_{zz} + C_{zzxz} (\omega) \, u_{xz}
\nonumber \\
& - \lambda \, S (1-S^2) K_{xx}(\nabla ) \, n_x - \lambda \, S (1-S^2) K_{xy}(\nabla )  \big[  Q_y + S\, \eta_{Ayx} + \sqrt{1-S^2}\, \eta_{Ayz} \big] \, ,
\\
\sigma_{yx} (\omega) &=  \textstyle{\frac{1}{2}} \, \Delta [\alpha + \lambda\, S] Q_y + \textstyle{\frac{1}{2}} \big\{ \alpha \, \Delta  [\alpha + \lambda\, S] + \bar{C} \cos^2 \theta + \nu_{xyxy} \, \partial_t  \big\} u_{xy}
\nonumber \\
&+ \textstyle{\frac{1}{2}} \big\{ \beta \, \Delta  [\alpha + \lambda\, S] + \bar{C} \sin \theta \, \cos \theta  + \nu_{yzyz} \, \partial_t  \big\} u_{yz}  \, ,
\\
\sigma_{yz} (\omega) &=   \textstyle{\frac{1}{2}} \, \Delta \big[\beta + \lambda\, \sqrt{1-S^2} \big] Q_y + \textstyle{\frac{1}{2}} \big\{ \alpha \, \Delta  \big[\beta + \lambda\, \sqrt{1-S^2} \big] + \bar{C} \sin \theta \, \cos \theta  + \nu_{yzyz} \, \partial_t  \big\} u_{xy}
\nonumber \\
&+ \textstyle{\frac{1}{2}} \big\{ \beta \, \Delta  \big[\beta + \lambda\, \sqrt{1-S^2} \big] + \bar{C} \sin^2 \theta + \nu_{yzyz} \, \partial_t  \big\} u_{yz}\, ,
\\
\sigma_{xy} (\omega) &=  \sigma_{yx} (\omega) +  \textstyle{\frac{1}{2}} \, (\lambda -1) S K_{yy}(\nabla )  \big\{  Q_y + S\, \eta_{Ayx} + \sqrt{1-S^2}\, \eta_{Ayz} \big\}  + \textstyle{\frac{1}{2}} \, (\lambda -1) S K_{yx}(\nabla )  \, n_x ,
\\
\sigma_{xz} (\omega) &=  \textstyle{\frac{1}{2}} \, C_{xxxz} (\omega) \, u_{xx} + \textstyle{\frac{1}{2}} \, C_{yyxz} (\omega) \, u_{yy}  + \textstyle{\frac{1}{2}} \, C_{zzxz} (\omega)\, u_{zz} + \textstyle{\frac{1}{2}} \, C_{xzxz} (\omega) \, u_{xz} 
\nonumber \\
& + \big[ \textstyle{\frac{1}{2}} \, (\lambda +1) - \lambda \, S^2  \big] \sqrt{1-S^2} \, \big\{ K_{xx}(\nabla )  \, n_x +  K_{xy}(\nabla )  \big[  Q_y + S\, \eta_{Ayx} + \sqrt{1-S^2}\, \eta_{Ayz} \big] \big\} \, ,
\\
\sigma_{zx} (\omega) &=  \textstyle{\frac{1}{2}} \, C_{xxxz} (\omega) \, u_{xx} + \textstyle{\frac{1}{2}} \, C_{yyxz} (\omega) \, u_{yy}  + \textstyle{\frac{1}{2}} \, C_{zzxz} (\omega)\, u_{zz} + \textstyle{\frac{1}{2}} \, C_{xzxz} (\omega) \, u_{xz} 
\nonumber \\
& + \big[ \textstyle{\frac{1}{2}} \, (\lambda - 1) - \lambda \, S^2  \big] \sqrt{1-S^2} \, \big\{ K_{xx}(\nabla )  \, n_x +  K_{xy}(\nabla )  \big[  Q_y + S\, \eta_{Ayx} + \sqrt{1-S^2}\, \eta_{Ayz} \big] \big\} \, ,
\\
\sigma_{zy} (\omega) &=   \sigma_{zy} (\omega) \, ,
\end{align} 
\end{subequations}
with $C_{ijkl} (\omega)$ as defined in Eq.~(\ref{defFreqDepMod}).
\end{widetext}

Since these equations of motion are of considerable algebraic complexity, we will first consider the simplified case where all the $K_{ijkl}$ are set to zero. This will allow us without too much effort to make contact to the hydrodynamic equations derived in Sec.~\ref{SmChydrodynamicEqs} and to extract renormalized elastic moduli that will be important for the behavior of Sm$C$ elastomers in rheology experiments. We will return to the full equations with the $K_{ijkl}$ included further below.

When the $K_{ijkl}$ vanish, the dependence of the stress tensor and Eq.~(\ref{QyEquation}) on $n_x$ drops out. The latter equation is then readily solved with the result
\begin{align}
\label{QyFourier}
Q_y &= - \alpha \, \frac{1 + i \omega \tau_3}{1 - i \omega \tau_1}
\,  u_{xy} - \beta \, \frac{1 + i \omega \tau_2}{1 - i \omega
\tau_1} \,  u_{yz} \,
\end{align}
where we have introduced the relaxation times 
\begin{align}
\tau_1 = 1/( \Gamma
\Delta) , 
\end{align}
$\tau_2 = \lambda \sqrt{1-S^2}/(\Gamma \Delta \beta)$,
$\tau_3 = \lambda S/( \Gamma \Delta \alpha)$. As we will see
further below, our dynamical equations will predict non-hydrodynamic
modes, which are essentially director modes, with a decay time (``mass") $\tau_1$ implying that $\tau_1$ is essentially the director relaxation time, $\tau_1 =
\tau_n$. To estimate the value of $\tau_n$, we note that the elastic constant $\Delta$  can be re-expressed in terms of the parameters of semi-microscopic models such as the neo-classical model of rubber elasticity~\cite{WarnerTer2003}. For details on the relation between the elastic constants of our elastic energy~(\ref{formOfF}) and those of the underlying semi-microscopic models we refer to Ref.~\cite{stenull_lubensky_SmCsoftness_2006}. Then, we use the known values of the semi-microscopic parameters to estimate the magnitude of $\Delta$. We find $\Delta$ to be of the order of $10^3 \, \mbox{Pa}$. We are not aware of any experimental or theoretical estimates of the rotational viscosity $\Gamma^{-1}$ of Sm$C$ elastomers. We expect it to be somewhat larger than the values of about $0.3 \times 10^{-1} \, \mbox{Pa} \, \mbox{s}$ found in Sm$C$ liquid crystals~\cite{meiboom_hewitt_1975}, perhaps $\Gamma^{-1} \approx 1 \,  \mbox{Pa} \, \mbox{s}$. This leads to $\tau_n \approx 10^{-3} \mbox{s}$ as a rough estimate of the director relaxation time in Sm$C$ elastomers. This value is larger than the  $\tau_n \approx 10^{-2} \, \mbox{s}$ reported for nematic elastomers~\cite{schmidtke&CO_2000,schornstein&Co_2001}, which is consistent with the expectation that restoring forces on the director are greater in smectics than in nematics due to the layering.

Feeding Eq.~(\ref{QyFourier}) into the stress tensor~(\ref{SmCStressTensStrainAndDir}) leads an effective stress tensor and corresponding effective equations of motion in terms of the displacement variables only. This effective stress tensor has exactly the same form as the hydrodynamic stress tensor~(\ref{SmCStressTensStrainOnly}) without the bending terms and with the renormalized frequency-dependent elastic moduli now given by
\begin{align}
\label{renormalizedModuli}
C_{\xi \chi}^R (\omega) &=  C_{\xi \chi} - i \omega  \, \nu_{\xi
\chi}  - \frac{i\omega \tau_1}{1 - i\omega \tau_1} \,  \Delta
A_{\xi \chi}
\nonumber \\
&= C_{\xi \chi} - i \omega  \, \nu_{\xi \chi}^R + O(\omega^2),
\end{align}
with renormalized viscosities 
\begin{align}
\label{renormalizedVoscosities}
\nu_{\xi \chi}^R = \nu_{\xi \chi} + \Gamma^{-1} A_{\xi \chi} \, .
\end{align}
In Eqs.~(\ref{renormalizedModuli}) and (\ref{renormalizedVoscosities}) we have used our compact notation for index pairs established in Eq.~(\ref{renormalizedModuliHydrodyn}). The individual $A_{\xi \chi}$ are given by $A_{xyxy} = \alpha^2 (1+ \tau_3/\tau_1)^2$, $A_{xyzy} = \alpha \beta (1+ \tau_3/\tau_1)(1+ \tau_2/\tau_1) $ and $A_{zyzy} = \beta^2 (1+ \tau_2/\tau_1)^2$.  
Note from Eq.~(\ref{renormalizedModuli}) that we can identify the bare viscosities $\eta_{ijkl}$ and $\nu_{ijkl}$ as well as the renormalized viscosities $\eta_{\xi \chi}^R$ and $\nu_{\xi \chi}^R$, indicating the consistency of our approaches without and with director.

\subsection{Mode structure in the incompressible limit}
\label{SmCmodeStructIncompStrainDir}
Now turn to the equations of motion, Eqs.~(\ref{nxQy}) and (\ref{compactEOMs}) in conjunction with (\ref{SmCStressTensStrainAndDir}), with the bending terms included and considering the incompressible limit. As above, we will focus on the directions where $\brm{q} = (q_x, 0, q_z)$ and $\brm{q} = (0, q_y, 0)$, respectively. Some of the steps involved in solving these equations of motion are sketched in appenix~\ref{solEqmWithDir} . Contrary to the predictions of our purely hydrodynamical theories, the modes resulting here are of three rather than of two different types. As was the case in the hydrodynamics of biaxial smectics and Sm$C$'s, there are propagating modes with frequencies in the form of Eq.~(\ref{wp}) and there are diffusive modes with frequencies as stated in Eqs.~(\ref{wd}) and (\ref{wsf}). In addition to these modes, however, our theory with strain and director produces non-hydrodynamic modes whose frequencies remain finite in the limit of vanishing wavevector. Their frequencies are of the form
\begin{align}
\label{wm}
\omega_{m} = - i \, \tau_1^{-1} + i \, D_m \, q^2 .
\end{align}
These non-hydrodynamic modes with a zero-$q$ decay time $\tau_1$ are predominantly director relaxation modes, and $\tau_1$ is the director relaxation time $\tau_n$. In the following we will discuss the specifics of all three types of modes.

Let us begin by considering case that $\brm{q}$ lies in the
$xz$-plane. In this case the equation of motion for $u_y$
decouples from the equations of motion for $u_x$ and $u_z$. The $u_y$
equation produces a set  of transverse modes with $\brm{u}$
along  $\tilde{\brm{e}}_y$. There is a non-hydrodynamic mode with
\begin{align}
\label{wym}
D_{m,y}=   \frac{\Big[\sqrt{\nu_{xyxy}^R - \nu_{xyxy}} \,
\hat{q}_x + \sqrt{\nu_{zyzy}^R - \nu_{zyzy}} \, \hat{q}_z
\Big]^2}{4 \, \rho} .
\end{align}
Moreover, there are propagating modes with a sound velocity $C_y$ that is identical to the sound velocity $C_y$ resulting from our hydrodynamic theory of Sec.~\ref{SmChydrodynamics}, see Eq.~(\ref{CyHydro}), and with a diffusion constant
\begin{align}
\label{Dpy}
D_{p,y}  &= \frac{\nu_{xyxy}^R \, \hat{q}_x^2 + 2 \, \nu_{xyzy}^R\,  \hat{q}_x \hat{q}_z + \nu_{zyzy}^R\, \hat{q}_z^2}{8 \, \rho }
\, .
\end{align}
In the soft direction, i.~e.\ for $\brm{q} \parallel
\tilde{\brm{e}}_{z^\prime}$, these propagating modes become diffusive with
$D_{d,y} = D_{p,y}$ and
\begin{align}
\label{By}
B_y  = \frac{\bar{K}_1 \, \hat{q}_x^4  + \bar{K}_2 \, 
\hat{q}_z^4 + \bar{K}_3 \, \hat{q}_x^2 \hat{q}_z^2 + 2 \, \bar{K}_4 \, 
\hat{q}_x^3 \hat{q}_z + 2 \,  \bar{K}_5 \, \hat{q}_x \hat{q}_z^3}{\rho} \, ,
\end{align}
where the $\bar{K}$'s are bending moduli that are combinations of
the Frank elastic constants, the order parameter $S$ as well
as $\lambda$, $\alpha$, and $\beta$. Our results for these $\bar{K}$'s are compiled in appendix~\ref{specificsBendingModIncompStrainDir}. We would like to emphasize, that the hydrodynamic theory and the theory with strain and director yield identical results for the sound velocity $C_y$. Furthermore, Eqs.~(\ref{Dpy}) and (\ref{DpyHydro}) as well as Eqs.~(\ref{By}) and (\ref{ByHydro}) show that the results for the diffusion constants $D_{p,y}$ and the bending terms $B_y$ are in absolute agreement, when the identifications $\nu_{xyxy}^R = \eta_{xyxy}^R$ and so on as well as $\bar{K}_1 = B_1$ and so on are made.

The equations of motion for $u_x$ and $u_z$ are solved by decomposing $(u_x,u_z)$ into a
longitudinal part $u_l$,  that vanishes in the incompressible limit,  and a transversal part
$u_T$. The  equation of motion for $u_T$ yields  a pair of propagating modes with a sound velocity as given in Eq.~(\ref{CThydro}), with 
\begin{align}
\label{DpT}
D_{p,T} &=  \frac{\nu_{xxxx} + \nu_{zzzz} - 2\, 
\nu_{xxzz}}{2\, \rho} \,  \hat{q}_x^2 \hat{q}_z^2
 \\
&+ \frac{ \nu_{xxxz}  -  \nu_{zzxz}}{2\, \rho} \,  \hat{q}_x \hat{q}_z (\hat{q}_z^2
-\hat{q}_x^2) + \frac{\nu_{xzxz}}{8\, \rho} \, (\hat{q}_z^2
-\hat{q}_x^2)^2 , \nonumber
\end{align}
and with a $\brm{u} \parallel \tilde{\brm{e}}_T$ where $\tilde{\brm{e}}_T = (\hat{q}_z,
0, -\hat{q}_x)$. Note the full agreement of Eqs.~(\ref{DpT}) with Eq.~(\ref{DpThydro}) when $\nu_{ijkl}$ and $\eta_{ijkl}$ are identified.

Finally, we turn to $\brm{q} \parallel \tilde{\brm{e}}_y$. In this case here
is a pair of longitudinal propagating modes with $\brm{u}
\parallel \tilde{\brm{e}}_y$ that is suppressed in the incompressible
limit. There is a non-hydrodynamic mode with
\begin{align}
\label{wxzm}
D_{m,xz}= \frac{ \nu_{xyxy}^R - \nu_{xyxy}+
\nu_{zyzy}^R - \nu_{zyzy}}{4\, \rho}  \,  ,
\end{align}
where $\brm{u}$ lays in the $xz$-plane with $u_x = \alpha (\tau_1
+\tau_3)/[\beta (\tau_1 +\tau_2)] \, u_z$. There is a pair of diffusive modes with polarization $\brm{u}
\parallel \tilde{\brm{e}}_{z^\prime}$ that is related to elastically soft deformations. The diffusion constant and the bending term of this mode pair reads
\begin{subequations}
\begin{align}
D_{d,z^\prime}  &= \frac{ \nu_{z^\prime y^\prime
z^\prime y^\prime}^R}{8 \, \rho}  \, ,
\\
B_{z^\prime}  &= \frac{\sin^2 \theta \,  \bar{K}_6 -
\sin 2 \theta \,\bar{K}_8   + \cos^2 \theta \, \bar{K}_7 }{\rho} \, ,
\end{align}
\end{subequations}
where $ \nu_{z^\prime y^\prime z^\prime y^\prime}^R$ is the
renormalized version of $ \nu_{z^\prime y^\prime z^\prime
y^\prime}$, cf.\ Eq.~(\ref{ViscositiesPrime}). Last, there is a pair of propagating modes polarized
along $\tilde{\brm{e}}_{x^\prime}$ with a sound velocity identical to that stated in Eq.~(\ref{CxPrimeHydro}) and with a diffusion constant
\begin{align}
D_{p,x^\prime}  &=   \frac{ \nu_{x^\prime y^\prime x^\prime y^\prime }^R}{8\, \rho}  \,  ,
\end{align}
where $\nu_{x^\prime y^\prime x^\prime y^\prime }^R = \cos^2 \theta
\,  \nu_{xyxy}^R + \sin 2 \theta \, \nu_{xyzy}^R  + \sin^2 \theta
\, \nu_{zyzy}^R$. Once more, we point out the full agreement of the diffusion constants and the bending terms obtained respectively through our purely hydrodynamical theory and our theory with strain and director when the proper identifications between the viscosities and between $\bar{K}_6$ and $B_6$ etc.\ are made. This agreement on one hand signals the consistency of our two approaches. On the other hand, it reassures us that our algebra and our final results are correct.

\subsection{Rheology}
\label{SmCrheology}
In the hydrodynamic limit $\omega \tau \to 0$, physical elastic moduli have the simple Maxwell (rubber-like) frequency dependence, i.e., $C_{ijkl} (\omega)= C_{ijkl} - i \omega \, \nu_{ijkl}$. At higher frequencies, they can exhibit more complex behavior arising from non-hydrodynamic modes. Our simple theory focuses on one set of non-hydrodynamic modes, those associated with the director. It provides a good description of the relaxation of the director, but it provides a good description of the dynamic elastic moduli only if $\tau_n$ is much greater than than any other relaxation time in the system, for example those associated with the relaxation of polymer configurations. If we assume $\tau_n \gg \tau_E$, then our theory makes specific predictions about the frequency dependent elastic moduli and the corresponding experimentally relevant quatities, viz.\ the storage and loss moduli defined, respectively, as the real and imaginary parts of the frequency dependent moduli.

\begin{figure}
\includegraphics[width=7.0cm]{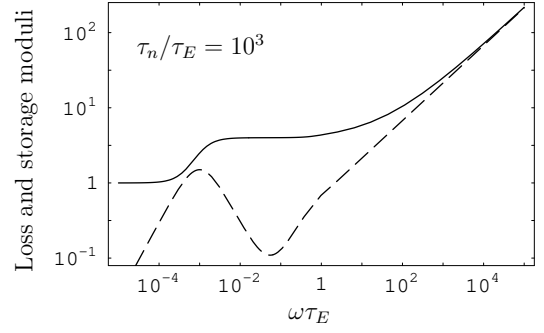}
\caption[]{\label{LossStorageModuli}Log-log plot of the reduced storage
and loss moduli $C_{\xi \chi}^\prime (\omega)/C_{\xi \chi}$ (solid lines) and
$C_{\xi \chi}^{\prime \prime}(\omega)/C_{\xi \chi}$ (dashed lines) versus the
reduced frequency $\omega \tau_E$ as given respectively by
the real and imaginary parts of Eq.~(\ref{complexModulusPhenoForm})
for $\tau_n/\tau_E=10^3$. For the purpose of illustration we have set, by and large
arbitrarily, $\nu_{\xi \chi} /(\tau_E C_{\xi \chi}) = 1$ and $\Delta A_{\xi \chi} /C_{\xi \chi} = 3$.}
\end{figure}

One consequence of Eq.~(\ref{renormalizedModuli}) is is that the corresponding storage and loss moduli may exhibit unconventional dips and plateaus when they are measured as functions of $\omega$. Because rheology experiments typically probe a wide range of frequencies that reaches to frequencies well above the hydrodynamic regime, it seems useful to extend our discussion to frequencies exceeding $\omega \ll \tau_E^{-1}$. In oder to do so, we have to account for the frequency dependence of the viscosities and $\Gamma$. The Rouse model~\cite{rubinstein_colby_2003}, though it certainly does not provide a microscopically correct description of liquid-crystal elastomer dynamics, is known to provide for useful fitting curves for the storage and loss moduli of elastomers. This observation leads us employ this model here and assume that the viscosities and $\Gamma$ are respectively of the form $\nu_{ijkl} (\omega)= \nu_{ijkl} f_R ( - i \omega \tau_E)$ and $\Gamma (\omega)^{-1}= \Gamma^{-1} f_R ( - i \omega \tau_E)$ where 
\begin{align}
\label{RousefunctionAna}
f_R ( x ) = \frac{3}{\pi^2  x  \sinh (\pi \sqrt{x})} \left\{
\pi \sqrt{x}  \cosh (\pi \sqrt{x}) -  \sinh (\pi \sqrt{x})
\right\}  ,
\end{align}
is the well known Rouse function~\cite{rubinstein_colby_2003}. With these assumptions, we obtain the following phenomenological form for the complex moduli $C_{\xi \chi}^R (\omega)$ that incorporates the effects of both director and elastomer modes: 
\begin{align}
\label{complexModulusPhenoForm}
C_{\xi \chi}^R (\omega) &=  C_{\xi \chi} +  \frac{\nu_{\xi \chi}}{\tau_E}   \, h_R ( - i \omega \tau_E)
\nonumber  \\
&+ \frac{(\tau_n/\tau_E) \, h_R ( - i \omega \tau_E)}{1 + (\tau_n/\tau_E) \, h_R ( - i \omega \tau_E)} \, \Delta A_{\xi \chi}  \, ,
\end{align}
where $h_R ( x ) = x\, f_R ( x )$. For  $\tau_n \gg \tau_E$, the storage moduli $C_{\xi \chi}^\prime$ produced by Eq.~(\ref{complexModulusPhenoForm}) show an unconventional plateau and there is an unusual dip in the loss moduli $C_{\xi \chi}^{\prime \prime}$, cf.\ Fig.~\ref{LossStorageModuli}.
If $\tau_n \approx \tau_E$, or $\tau_n < \tau_E$, this is not the case because elastomer modes will blur these plateaus and dips leading to conventional rubber-like behavior (not shown).

There is another interesting theoretical implication of Eq.~(\ref{renormalizedModuli}) that, however, will be very difficult to detect experimentally. To discuss this implication, let us switch here to the rotated reference space coordinates $x^\prime, y^\prime, z^\prime$. Experimentally, it will be hard to work with these coordinates, since the corresponding rotation angle $\theta$ depends on temperature. Setting this problem aside, we note that there are two renormalized shear moduli that vanish in the limit $\omega \to 0$, viz.\ 
\begin{subequations}
\label{renormalizedModuliPrime}
\begin{align}
C_{y^\prime z^\prime y^\prime z^\prime}^R (\omega) &=   - i \omega
\, \nu_{y^\prime z^\prime y^\prime z^\prime}  - \frac{i\omega
\tau_1}{1 - i\omega \tau_1} \,  \Delta A_{y^\prime z^\prime
y^\prime z^\prime} \, ,
\\
C_{x^\prime y^\prime y^\prime z^\prime}^R (\omega) &=   - i \omega
\, \nu_{x^\prime y^\prime y^\prime z^\prime}  - \frac{i\omega
\tau_1}{1 - i\omega \tau_1} \,  \Delta A_{x^\prime y^\prime
y^\prime z^\prime} \, ,
\end{align}
\end{subequations}
with viscosities
\begin{subequations}
\label{ViscositiesPrime}
\begin{align}
\nu_{y^\prime z^\prime y^\prime z^\prime} &=  \sin^2 \theta \, 
\nu_{xyxy} - \sin 2 \theta \, \nu_{xyyz} + \cos^2 \theta \,  \nu_{yzyz} \, ,
\\
\nu_{x^\prime y^\prime y^\prime z^\prime} &=   \cos 2\theta \, \nu_{xyyz} + \textstyle{\frac{1}{2}} \, \sin 2 \theta \, (\nu_{yzyz} - \nu_{xyxy})\, .
\end{align}
\end{subequations}
 $A_{y^\prime z^\prime y^\prime z^\prime}$ and $A_{x^\prime y^\prime y^\prime z^\prime}$ can be inferred from Eqs.~(\ref{ViscositiesPrime}) upon replacing $\nu_{xyxy}$ by $A_{xyxy}$ and so on. Since $C_{y^\prime z^\prime y^\prime z^\prime}^R (\omega)$ and
$C_{x^\prime y^\prime y^\prime z^\prime}^R (\omega)$ vanish in the for  $\omega \to 0$, we recover in this limit the ideal static soft elasticity discussed in Sec.~\ref{SmCelasticEn}.  At non-vanishing frequency the system cannot be ideally soft but it can be nearly so for $\omega$ small. This deviation from ideal softness due to non-zero frequencies was first discussed in the context of nematic elastomers~\cite{terentjev&Co_NEhydrodyn}, where this phenomenon was termed dynamic soft elasticity. In a semi-soft Sm$C$, the storage moduli
$C_{y^\prime z^\prime y^\prime z^\prime}^\prime (\omega)$ and
$C_{x^\prime y^\prime y^\prime z^\prime}^\prime (\omega)$ in the
rotated frame are nonzero at zero frequency. They will exhibit behavior similar to that
Fig.~\ref{LossStorageModuli} if $\tau_n \gg \tau_E$. 

\section{Concluding remarks}
\label{concludingRemarks}
In summary, we have studied the low-frequency, long-wavelength dynamics of Sm$A$, biaxial smectic and Sm$C$ elastomers, assuming that these materials have been crosslinked in the Sm$A$ phase. We employed two different but related approaches: one formulation that does not explicitly involve the Frank director and that describes pure hydrodynamics and a second formulation that features the director and that describes slow modes beyond the hydrodynamic limit.

The hydrodynamics of Sm$A$ elastomers is qualitatively the same as that of conventional uniaxial rubbers. Beyond the hydrodynamic regime, however, the director has an impact on the dynamics of Sm$A$ elastomers, in which the genuine difference between Sm$A$ and conventional uniaxial elastomers shows up. The low-frequency, long-wavelength dynamics of Sm$A$ elastomers, as described by our theory with the director, is qualitatively the same as that of nematic elastomers, up to those aspects, where the value of the modulus $C_5^R$ for shears in the planes containing the director enters. For example, nematic and Sm$A$ elastomers possess 3 pairs of hydrodynamic sound modes. One of the qualitative differences is, that a soft nematic elastomer, where $C_5^R =0$, has two pairs of sound modes whose sound velocity vanishes in the softness related symmetry directions, whereas for a Sm$A$ elastomers, where $C_5^R$ is significantly larger than zero, sound velocities are non-vanishing in all directions. The other qualitative difference is that nematic elastomers exhibit dynamic (semi-)soft elasticity, i.e., the generalization of (semi-)soft elasticity to non-zero frequencies. Sm$A$ elastomers crosslinked in the Sm$A$ phase, since they are not soft or semi-soft, cannot have this property.

Biaxial smectic and Sm$C$ elastomers possess, like nematic and Sm$A$ elastomers, three pairs of hydrodynamic sound modes. One of the sound-mode pairs in either elastomer is associated with soft deformations. The sound velocity of these pairs, which become purely transverse in the incompressible limit, vanishes in the softness-related symmetry directions. The remaining two sound sound mode pairs are associated with non-soft deformations. Their velocities remain finite in the directions in which the velocities of the softness-related modes vanishes. In the incompressible limit, these mode pairs become purely transverse and longitudinal, and hence the latter are effectively suppressed. The vanishing of the velocities of the softness-related modes in certain directions could be exploited, in principle, to separate these modes from the remaining modes, which have different transverse polarizations than the softness-related modes. Therefore, at least theoretically, soft biaxial smectic and Sm$C$ elastomers could be used to polarize acoustic waves.

Our theory with strain and director reveals that Sm$C$ elastomers may have an unconventional behavior in rheology experiments. It predicts that, as in nematic and Sm$A$ elastomers, certain storage and loss moduli may show, respectively, a plateau and an associated dip in the frequency range $\tau_n^{-1} < \omega < \tau_E^{-1}$. It also predicts that soft samples of these materials exhibit dynamic soft elasticity. Recently there has been a controversy whether this unconventional rheology and dynamic soft elasticity can be observed in liquid crystal elastomers or not. Thus far, this controversy revolved around nematic elastomers. The few samples of nematic elastomers on which rheology measurements have been performed, produced little evidence for unconventional rheology or dynamic soft elasticity. However, due to the limited amount of data that is currently available and due to differences in their interpretation, the experimental picture still gives reason for debate; both phenomena are neither clearly verified nor clearly ruled out. The idea to use Sm$C$ elastomers as an alternative testing ground for these phenomena seems appealing. As far as dynamic soft or semi-soft elasticity is concerend, however, it will be difficult if not impossible to realize soft or semi-soft oszillatory shears experimentally, because one has to use a very specific, temperature-dependent coordinate system. Regarding the plateau and the dip in, respectively, the storage and loss moduli, these are less likely to exist physically in smectics than in nematics, because the director relaxation time in smectic elastomers is probably shorter than in nematic elastomers. 

Despite these difficulties, we hope that our theory encourages rheology experiments on smectic elastomers, in particular, to see if unconventional rheology exist or not. Moreover, we hope that our work motivates experimental investigations of sound velocities in smectic elastomers, for example Brillouin scattering experiments.

\acknowledgements
Support by the National Science Foundation under grant DMR 0404670 (TCL) is gratefully acknowledged. 

\appendix

\section{Elastic moduli}
This Appendix collects specifics about the Frank elastic constants defined by Eq.~(\ref{FrankExpanded}), the bending moduli defined by Eq.~(\ref{BendingEn}) and the effective bending moduli appearing in Sec.~\ref{SmCmodeStructIncompStrainDir}.  

\subsection{Frank elastic constants appearing in Eq.~(\ref{FrankExpanded})}
\begin{subequations}
\label{specificsFrankConst}
\begin{align}
K_{xxxx} &= K_1 + K_3 \, \frac{S^2}{1 - S^2}\, ,
\\
K_{yyyy} &= K_1 \, ,
\\
K_{yxyx} &= K_2 \, (1-S^2) + K_3 \, S^2 \, ,
\\
K_{xyxy} &= K_2 \, \frac{1}{1 - S^2}  \, ,
\\
K_{xzxz} &= K_1 \, \frac{S^2}{1 - S^2} + K_3 \, ,
\\
K_{yzyz} &= K_2 \, S^2 + K_3  \, (1-S^2) \, ,
\\
K_{xxyy} &= K_1 \, ,
\\
K_{xyyx} &= K_2 \, ,
\\
K_{xxxz} &= [ K_3 - K_1] \, \frac{S}{\sqrt{1 - S^2}}  \, ,
\\
K_{yyxz} &=  - K_1 \, \frac{S}{\sqrt{1 - S^2}}  \, ,
\\
K_{yxyz} &= [ K_3 - K_2] \, S \, \sqrt{1 - S^2}  \, ,
\\
K_{xyyz} &=  K_2 \, \frac{S}{\sqrt{1 - S^2}}  \, .
\end{align}
\end{subequations}

\subsection{Bending moduli appearing in Eq.~(\ref{BendingEn})}
\begin{subequations}
\label{specificsBendingMod}
\begin{align}
B_1 &= \textstyle{\frac{1}{4}} \, K_{yxyx} \, (S-\alpha)^2 ,
\\
B_2 &= \textstyle{\frac{1}{4}} \, K_{yzyz} \, (\sqrt{1 - S^2} - \beta)^2  ,
\\
B_3 &=  \textstyle{\frac{1}{4}} \, \big\{  K_{yxyx} \, (\sqrt{1 - S^2} - \beta)^2 + K_{yzyz} \, (S-\alpha)^2 
\nonumber \\
&+ 4 \, K_{yxyz} \, (S-\alpha) (\sqrt{1 - S^2} - \beta) \big\}  ,
\\
B_4 &= \textstyle{\frac{1}{4}} \, \big\{  K_{yxyx} \, (S-\alpha) (\sqrt{1 - S^2} - \beta) 
\nonumber \\
&+ K_{yxyz} \, (S-\alpha)^2 \big\}  ,
\\
B_5 &= \textstyle{\frac{1}{4}} \, \big\{  K_{yzyz} \, (S-\alpha) (\sqrt{1 - S^2} - \beta) 
\nonumber \\
&+ K_{yxyz} \, (\sqrt{1 - S^2} - \beta)^2 \big\}  ,
\\
B_6 &= \textstyle{\frac{1}{4}} \, K_{yyyy} \, (S+\alpha)^2 ,
\\
B_7 &= \textstyle{\frac{1}{4}} \, K_{yyyy} \, (\sqrt{1 - S^2} + \beta)^2  ,
\\
B_8 &= \textstyle{\frac{1}{4}} \, K_{yyyy} \, (S+\alpha) (\sqrt{1 - S^2} + \beta) \,  .
\end{align}
\end{subequations}

\subsection{Bending moduli appearing in Sec.~\ref{SmCmodeStructIncompStrainDir}}
\label{specificsBendingModIncompStrainDir}
\begin{subequations}
\begin{align}
\bar{K}_1 &=  \textstyle{\frac{1}{4}} \, K_{yxyx} \, (\alpha + \lambda S) (\alpha - S),
\\
\bar{K}_2 &=  \textstyle{\frac{1}{4}} \, K_{yzyz} \, (\beta + \lambda \sqrt{1 - S^2}) (\beta - \sqrt{1 - S^2}) ,
\\
\bar{K}_3 &=  \textstyle{\frac{1}{4}} \, \big\{  K_{yxyx} \,  (\beta + \lambda \sqrt{1 - S^2}) (\beta - \sqrt{1 - S^2}) 
\nonumber \\
&+ K_{yzyz} \, (\alpha + \lambda S) (\alpha - S) 
\nonumber \\
&+ 2 \, K_{yxyz} \, [  (\alpha + \lambda S)   (\beta - \sqrt{1 - S^2}) 
\nonumber \\
&+ (\beta + \lambda \sqrt{1 - S^2}) (\alpha - S)  ]     \big\}  ,
\\ 
\bar{K}_4 &= \textstyle{\frac{1}{8}} \, \big\{  K_{yxyx} \,  [  (\alpha + \lambda S)   (\beta - \sqrt{1 - S^2}) 
\nonumber \\
&+ (\beta + \lambda \sqrt{1 - S^2}) (\alpha - S)  ]
\nonumber \\
&+ 2 \, K_{yxyz} \, (\alpha + \lambda S) (\alpha - S) \big\} ,
\\
\bar{K}_5 &=   \textstyle{\frac{1}{8}} \, \big\{  K_{yxyx} \,  [  (\alpha + \lambda S)   (\beta - \sqrt{1 - S^2}) 
\nonumber \\
&+ (\beta + \lambda \sqrt{1 - S^2}) (\alpha - S)  ]
\nonumber \\
&+ 2 \, K_{yxyz} \, (\beta + \lambda \sqrt{1 - S^2}) (\beta - \sqrt{1 - S^2}) \big\} ,
\\
\bar{K}_6 &=  K_{yyyy} \,  \big\{  (\alpha + \lambda S) (\alpha + S)  
\nonumber \\
&- S (\lambda -1)(\alpha - S) \big\}  ,
\\
\bar{K}_7 &=  K_{yyyy} \, (\beta + \lambda \sqrt{1 - S^2}) (\beta + \sqrt{1 - S^2}) ,
\\
\bar{K}_8 &=  K_{yyyy} \, \big\{  2\beta(\alpha + S)
\nonumber \\
&+ (\alpha + 3 \lambda S - \alpha \lambda -S)  \sqrt{1 - S^2}  \big\}   .
\end{align}
\end{subequations}

\section{Solving the equations of motion in the Sm$C$ dynamics with strain and director}
\label{solEqmWithDir}
Here we sketch our analysis of the equations of motion, Eqs.~(\ref{nxQy}) and (\ref{compactEOMs}) together with (\ref{SmCStressTensStrainAndDir}), in order to obtain the modes discussed in Sec.~\ref{SmCmodeStructIncompStrainDir}.

\begin{widetext}
Let us start be introducing the Fourier transforms of the differential operators $K_{xx}(\nabla )$ and so on defined in Eq.~(\ref{KxxAndSoOn}):
\begin{subequations}
\begin{align}
q^2 K_{xx} &\equiv K_{xxxx} \, q_x^2 + K_{xyxy} \, q_y^2 + K_{xzxz} \, q_z^2 
+ 2\, K_{xxxz} \, q_x q_z \, ,
\\
q^2 K_{yy} &\equiv  K_{yxyx} \, q_x^2 + K_{yyyy} \, q_y^2 + K_{yzyz} \, q_z^2 
+ 2\, K_{yxyz} \, q_x q_z \, ,
\\
q^2 K_{xy} &\equiv  q_y \big\{  (K_{xxyy} + K_{xyyx})  \, q_x 
+ (K_{yyxz} + K_{xyyz}) \,  q_z \big\} \, ,
\\
q^2 K_{yx} &\equiv q_y \big\{  (K_{xxyy} + K_{xyyx})  \, q_x 
+ (K_{yyxz} + K_{yxyz}) \, q_z \big\} \, .
\end{align}
\end{subequations}
When $q_y =0$ or $q_x = q_z =0$, the cross terms $q^2 K_{xy}$ and $q^2 K_{yx}$ vanish. Writing Eqs.~(\ref{nxQy}) in momentum space one observes,  that in this case, the equations for $Q_y$ and $n_x$, decouple. This decoupling simplifies the solution of the equations of motion considerably and we will limit our following consideration to momenta where this simplification applies. 

When Eqs.~(\ref{nxQy}) decouple, they are readily solved for given displacements $u_x$, $u_y$ and $u_z$ with the result
\begin{subequations}
\label{QynxRes}
\begin{align}
Q_y &= - \alpha \, \frac{1 + i \omega \tau_3 + \frac{S}{\Delta \alpha} \, q^2 K_{yy}}{1 - i \omega \tau_1 + \frac{1}{\Delta } \, q^2 K_{yy}} \, {\textstyle{\frac{1}{2}}}\,  i q_x u_y  - \alpha \, \frac{1 + i \omega \tau_3 - \frac{S}{\Delta \alpha} \, q^2 K_{yy}}{1 - i \omega \tau_1 + \frac{1}{\Delta } \, q^2 K_{yy}} \, \textstyle{\frac{1}{2}}\,  i q_y u_x
\nonumber \\
& - \beta \, \frac{1 + i \omega \tau_2 + \frac{\sqrt{1-S^2}}{\Delta \beta} \, q^2 K_{yy}}{1 - i \omega \tau_1 + \frac{1}{\Delta } \, q^2 K_{yy}} \, {\textstyle{\frac{1}{2}}}\,  i q_z u_y - \beta \, \frac{1 + i \omega \tau_2 - \frac{\sqrt{1-S^2}}{\Delta \beta} \, q^2 K_{yy}}{1 - i \omega \tau_1 + \frac{1}{\Delta } \, q^2 K_{yy}} \, \textstyle{\frac{1}{2}}\,  i q_y u_z \, ,
\\
n_x &= \left[  1 - \frac{q^2 K_{xx}}{i \omega \Gamma^{-1}} \right]^{-1} \Big\{ \lambda S (1-S^2) \, i q_x u_x  - \lambda S (1-S^2) \, i q_z u_z + \sqrt{1 -S^2} \big[ \lambda (1-2S^2) + 1 \big]  \, \textstyle{\frac{1}{2}}\,  i q_z u_x  
\nonumber \\
&+ \sqrt{1 -S^2} \big[ \lambda (1-2S^2) - 1 \big]  \, \textstyle{\frac{1}{2}}\,  i q_x u_z  \Big\} \, .
\end{align}
\end{subequations}
Substituting Eqs.~(\ref{QynxRes}) into the stress tensor~(\ref{SmCStressTensStrainAndDir}), one obtains effective equations of motion for the displacements only. 

As a specific example, let us now consider the case $q_y =0$ in some detail. The second case, $q_x = q_z =0$, can be treated by similar means and will be left as an exercise to the reader. For $q_y =0$, the effective equation of motion for $u_y$ decouples from those for $u_x$ and $u_z$, which remain coupled. Expanding in powers of $q$ and $\omega$, we find the equation of motion for $u_y$ to read
\begin{align}
\label{effectiveUyEq}
\rho \, \omega^2 \, u_y &= \Big\{ \textstyle{\frac{1}{4}} \, C^R_{xyxy} (\omega) \, q_x^2 + \textstyle{\frac{1}{4}} \, C^R_{yzyz} (\omega) \, q_z^2 + \textstyle{\frac{1}{4}} \, C^R_{xyyz} (\omega) \, q_x q_z 
\nonumber \\
& +\bar{K}_1  \, q_x^4 +  \bar{K}_2  \, q_z^4 + \bar{K}_3  \, q_x^2 q_z^2 + 2 \, \bar{K}_4  \, q_x^3 q_z + 2 \, \bar{K}_5  \, q_x q_z^3  \Big\} \, u_y   \, ,
\end{align}
where we have dropped higher order terms that do not affect our results. The bending moduli $\bar{K}_1$ and so on are as defined in App.~\ref{specificsBendingModIncompStrainDir}. Solving Eq.~(\ref{effectiveUyEq}) for frequencies in the form of a power series in $q$ then leads to the non-hydrodynamic mode with the diffusion constant stated in Eq.~(\ref{wym}) and the hydrodynamic modes with sound velocities, diffusion constants and bending contributions as given in Eqs.~(\ref{CyHydro}),  (\ref{Dpy}) and (\ref{By}). 

Finally, let us look at the coupled equations for $u_x$ and $u_z$. Proceeding like above, we obtain
\begin{subequations}
\label{effectiveUxUzEq}
\begin{align}
\rho \, \omega^2 \, u_x &= \Big\{ C_{xxxx} (\omega) \, q_x^2 +  \textstyle{\frac{1}{4}} \,  C_{xzxz} (\omega) \, q_z^2  + C_{xxxz} (\omega) \, q_x q_z \Big\} \, u_x  
\nonumber \\
&+ \Big\{ \textstyle{\frac{1}{2}} \, C_{xxxz} (\omega) \, q_x^2 + \textstyle{\frac{1}{2}} \, C_{zzxz} (\omega) \, q_z^2 + \big[  C_{xxzz} (\omega) +  \textstyle{\frac{1}{4}} \,  C_{xzxz} (\omega) \big]  q_x q_z\Big\} \, u_z 
 \\
\rho \, \omega^2 \, u_z &= \Big\{ \textstyle{\frac{1}{2}} \, C_{xxxz} (\omega) \, q_x^2 + \textstyle{\frac{1}{2}} \, C_{zzxz} (\omega) \, q_z^2 + \big[  C_{xxzz} (\omega) +  \textstyle{\frac{1}{4}} \,  C_{xzxz} (\omega) \big]  q_x q_z \Big\} \, u_x  
\nonumber \\
&+ \Big\{ \textstyle{\frac{1}{4}} \,  C_{xzxz} (\omega) \, q_x^2 +  C_{zzzz} (\omega) \, q_z^2  + C_{zzxz} (\omega) \, q_x q_z  \Big\} \, u_z \, ,
\end{align}
\end{subequations}
as the leading contributions. Equations~(\ref{effectiveUxUzEq}) can be decoupled in the incompressible limit. To this end, we switch now from the coordinate system with basis $\{ \tilde{\brm{e}}_x, \tilde{\brm{e}}_y, \tilde{\brm{e}}_z \}$ to a rotated system with basis $\{ \hat{q}, \tilde{\brm{e}}_y, \tilde{\brm{e}}_T \}$ with $\tilde{\brm{e}}_T  = (\hat{q}_z, 0, - \hat{q}_x)$. In the incompressible limit the longitudinal component $u_l$ of the displacement vanishes. The equation of motion for the transversal component $u_T$ along  $\tilde{\brm{e}}_T$ follows from Eqs.~(\ref{effectiveUxUzEq}) as
\begin{align}
\label{effectiveUTEq}
\rho \, \omega^2 \, u_T &= q^{-2} \Big\{   {\textstyle{\frac{1}{4}}} \,  C_{xzxz} (\omega) \, (q_z^2 - q_x^2)^2 + \big[  C_{xxxz} (\omega) - C_{zzxz} (\omega) \big] \,  q_x q_z (q_z^2 - q_x^2)
\nonumber \\
&+  \big[  C_{xxxx} (\omega) - 2\, C_{xxzz} (\omega) + C_{zzzz} (\omega)  \big] \, q_x^2 q_z^2  \Big\} \, u_T   \, .
\end{align}
Solving this equation for $\omega$ in the form of a power series in $q$ results in the propagating modes with a sound velocity and a diffusion constant as resented in Eqs.~(\ref{CThydro}) and (\ref{DpT}).
\end{widetext}


\begin{thebibliography}{}
\bibitem{footnoteDiffConst}
Here and in the following we use the notion "diffusion constant'' somewhat loosely.
\end{thebibliography}
\end{document}